\documentclass[structabstract]{aa} 
\usepackage{graphicx}
\usepackage{txfonts}
\usepackage{natbib}       

\def\msun{M$_{\odot}$}
\def\h{$^{\rm h}$}\def\m{$^{\rm m}$}

\def\degs{\ifmmode ^{\circ}\else$^{\circ}$\fi}
\def\fss{\hbox{$.\!\!^{\rm s}$}}        
\def\farcs{\hbox{$.\!\!^{\prime\prime}$}}  

\def\amin{\ifmmode ^{\prime}\else$^{\prime}$\fi}
\def\asec{\ifmmode ^{\prime\prime}\else$^{\prime\prime}$\fi}
\newbox\grsign \setbox\grsign=\hbox{$>$}
\newdimen\grdimen \grdimen=\ht\grsign
\newbox\laxbox \newbox\gaxbox
\setbox\gaxbox=\hbox{\raise.5ex\hbox{$>$}\llap
     {\lower.5ex\hbox{$\sim$}}}\ht1=\grdimen\dp1=0pt
\setbox\laxbox=\hbox{\raise.5ex\hbox{$<$}\llap
     {\lower.5ex\hbox{$\sim$}}}\ht2=\grdimen\dp2=0pt
\def\gax{$\mathrel{\copy\gaxbox}$}
\def\lax{$\mathrel{\copy\laxbox}$}
\newcommand{\epse}{\overline{\mskip-1.27 \thinmuskip \epsilon_e \mskip-0.4 \thinmuskip}}
\begin{document}
   \title{The unusual afterglow of the Gamma-Ray Burst 100621A\thanks{Based 
on data acquired with the Atacama Pathfinder
Experiment (APEX) under ESO programme 285.D-5035(A). }}

   \author{J. Greiner\inst{1} \and
           T. Kr\"uhler\inst{2} \and
           M. Nardini\inst{3} \and
           R. Filgas\inst{1,4}
              \and
           A. Moin\inst{5,6}
           \and
           C. de Breuck\inst{7} \and
           F. Montenegro-Montes\inst{7} \and
           A. Lundgren\inst{8} \and
           S. Klose\inst{9} \and
           P.M.J. Afonso\inst{10} \and
           F. Bertoldi\inst{11} \and
           J. Elliott\inst{1} \and
           D.A. Kann\inst{9} \and
           F. Knust\inst{1} \and
           K. Menten\inst{12} \and
           A. Nicuesa Guelbenzu\inst{9} \and
           F.\,\,Olivares\,E.\inst{1} \and
           A. Rau\inst{1} \and
           A. Rossi\inst{9} \and
           P. Schady\inst{1} \and
           S. Schmidl\inst{9} \and
           G. Siringo\inst{8} \and
           L. Spezzi\inst{13}
           V. Sudilovsky\inst{1} \and
           S.J. Tingay\inst{5} \and
           A.C. Updike\inst{14} \and
           Z. Wang\inst{6} \and
           A. Weiss\inst{12} \and
           M. Wieringa\inst{15} \and 
           F. Wyrowski\inst{12}
          }
   \institute{Max-Planck-Institut f\"ur extraterrestrische Physik,
              Giessenbachstrasse 1, 85748 Garching, Germany
         \and
             Dark Cosmology Centre, Niels Bohr Institute, University of Copenhagen, 
             Juliane Maries Vej 30, 2100 Copenhagen, Denmark 
         \and
             Universit\'{a} degli studi di Milano-Bicocca, Piazza della 
             Scienza 3, 20126 Milano, Italy
         \and
              Institute of Experimental and Applied Physics, 
              Czech Technical University Prague, Horska 3a/22,
              128\,00 Prague 2, Czech Republic
         \and
             International Centre for Radio Astronomy Research, 
             Curtin University, GPO Box U1987, Perth, WA 6845, Australia
         \and
             Shanghai Astronomical Observatory, Chinese Academy of Sciences 
             (SHAO), 80 Nandan Road,  Shanghai 200030, China
         \and
              ESO, Alonso de C\'{o}rdoba  3107, Vitacura, Casilla 19001, 
              Santiago, Chile
         \and
              ALMA JAO, Alonso de Cordova 3107, Vitacura, Casilla 19001,
              Santiago, Chile
         \and
             Th\"uringer Landessternwarte Tautenburg, Sternwarte 5,
             07778 Tautenburg,  Germany 
         \and
             American River College, Physics Dept., 4700 College Oak Drive,
             Sacramento, CA 95841, USA
         \and
             Argelander-Institut f\"ur Astronomie, Auf dem H\"ugel 71, 
             53121 Bonn, Germany
         \and
             Max-Planck-Institut f\"ur Radioastronomie, Auf dem H\"ugel 69,
             53121 Bonn, Germany
         \and
             European Southern Observatory, Schwarzschild-Str. 2,
             85748 Garching, Germany 
         \and
             Roger Williams Univ., One Old Ferry Road, Bristol, RI 02809, USA
         \and
             CSIRO Astronomy \& Space Science, Locked Bag 194, Narrabri,
             NSW 2390, Australia
             }

   \date{Received Feb 13, 2013; accepted Apr 17, 2013}

  \abstract
   {}
    {With the afterglow of GRB 100621A being the brightest detected so
     far in X-rays,
     and superb GROND coverage in the optical/NIR during the first
     few hours, an observational verification of basic fireball
     predictions seemed possible.
    }
   {In order to constrain the broad-band spectral energy distribution
    of the afterglow of GRB 100621A, dedicated
    observations were performed in the optical/near-infrared with the 7-channel 
    ``Gamma-Ray Burst Optical and Near-infrared Detector'' (GROND) at the 
    2.2m MPG/ESO telescope, in the sub-millimeter band with the large 
    bolometer array LABOCA at APEX, and at radio frequencies with ATCA. 
    Utilizing also Swift X-ray observations, we attempt an interpretation 
    of the observational data within the fireball scenario.
   }
    {The afterglow of GRB 100621A shows a very complex temporal as
     well as spectral evolution. We identify three different emission
     components, the most spectacular one causing a sudden intensity jump
     about one hour after the prompt emission. The spectrum of this 
     component is much steeper than the canonical afterglow. We interpret
     this component using the prescription of Vlasis et al. (2011)
     for a two-shell collision after the first shell has been decelerated
     by the circumburst medium. We use the fireball scenario to derive
     constraints on the microphysical parameters of the first shell.
     Long-term energy injection into a narrow jet seems to provide an
     adequate description.
     Another noteworthy result is the large ($A_V$ = 3.6 mag) 
     line-of-sight host extinction of the afterglow in an otherwise extremely
     blue host galaxy.
    }
    {Some GRB afterglows have shown complex features, and that of GRB 100621A
     is another good example. Yet, detailed observational campaigns of
     the brightest afterglows promise to deepen our understanding
     of the formation of afterglows and the subsequent interaction
     with the circumburst medium.
    }


   \maketitle
%

\section{Introduction}

\subsection{The fireball scenario}

Gamma-Ray Bursts (GRBs) are generally accepted to be related to the death of
massive stars. Due to their large gamma-ray luminosity, GRBs can be
detected to very high redshift, and thus provide a unique probe into the
Early Universe. Understanding the emission mechanism and geometry is 
crucial for deriving the burst energetics and number density, and 
observing and understanding the afterglow emission is of utmost   
importance to decipher the burst environmental properties (e.g., gas
density profile, metallicity, dust) as well as to derive constraints on
the progenitor (e.g., mass, rotation, binarity, supernova relation).

The late emission at X-ray to optical/radio wavelengths, the so-called 
afterglow, is dominated by synchrotron emission from the external shock, 
i.e. emission from relativistic electrons gyrating in a magnetic field
\citep{mer97, wij97, wig99}. 
This synchrotron shock model
is widely accepted as the major radiation mechanism in the external shock,
and the macroscopic properties of such shocks are largely understood. 
Under the implicit assumptions that the electrons are ``Fermi'' accelerated 
at the relativistic shocks to a power law
distribution with an index $p$ upon acceleration,
their dynamics can be expressed
in terms of 4 main parameters:
(1) the total internal energy in the shocked region as released in the 
explosion, 
(2) the electron density and radial profile of the  surrounding medium, 
(3)  fraction of the shock energy going into the ISM electrons
$\epsilon_e$,
(4) the fraction of energy density in the  magnetic field $\epsilon_B$. 
Measuring the energetics
and the energy partition ($\epsilon_e / \epsilon_B$) was possible 
only for a handful of the more than 900 GRB afterglows
so far, as it requires truly multi-wavelength observations between
X-rays and radio frequencies.
Moreover, there are large uncertainties in the microphysics: How are
the relativistic particles accelerated? How is the magnetic field
in the shocked region generated? What is its structure and evolution?
Addressing these questions is even more challenging.

According to standard synchrotron theory, the radiation
power of an electron with co-moving energy $\gamma_e m c^2$
 is $P_e = 4/3 \sigma_T c \gamma_e^2 (B^2/8\pi)$,
so that high energy electrons cool more rapidly.
For a continuous injection of electrons, which is the case for
ongoing plowing of the forward shock into the interstellar medium (ISM),
there is a break in the electron spectrum,
above which the spectrum is steepened due to cooling.
This energy is time-dependent, so this frequency break moves to lower
energies for the ISM case and opposite for a wind medium. 
Since the spectral slope as well as the temporal decay slope
are identical for the two density profiles, it is just the direction
of the cooling break movement that allows to distinguish between ISM and
wind density profile surrounding the GRB.

Besides this cooling frequency $\nu_c$, there is
the injection frequency $\nu_m$, corresponding to the electrons
accelerated in the shock to a power-law distribution with a minimum 
Lorentz factor, and the self-absorption frequency, $\nu_{sa}$.
The final GRB afterglow spectrum is thus a four-segment
broken power law 
\citep{mes98, sari98}
separated by $\nu_{sa}$, $\nu_m$, and $\nu_c$. 
The order of $\nu_m$ and $\nu_c$ defines two types
of spectra, namely the ``slow cooling case'' with $\nu_m < \nu_c$,
and the ``fast cooling case''  $\nu_m > \nu_c$.
For each case, and depending on wind vs. ISM density profile,
theory \citep{sari99} predicts different slopes of the power law
segments and speeds at which $\nu_m$ and $\nu_c$ should be moving.
For ``standard'' parameters, $\nu_m$ should be moving from 10$^{14}$ Hz
to 10$^{12}$ Hz within the first day, and $\nu_c$ from 10$^{17}$ Hz
to 10$^{14}$ Hz.
Due to sensitivity limitations
in the sub-mm range, and lack of coordinated multi-wavelength observations,
there is not a single GRB data set sufficient (in terms of wavelength
and temporal coverage)  to unambiguously verify these
predictions for both frequencies, and just two GRBs where the high-frequency
break (interpreted as cooling break) has been unambiguously
shown to move \citep{bbb06, fgs12}. 

\subsection{GROND and GRB 100621A}

GROND, a simultaneous 7-channel optical/near-infrared ima\-ger \citep{gbc08} 
moun\-ted at  the 2.2\,m telescope of the Max-Planck-Gesellschaft (MPG), 
operated by MPG and ESO (European Southern Observatory) at La Silla 
(Chile), started operation in May 2007. 
GROND has been built as a dedicated GRB follow-up 
instrument
and has observed basically every GRB visible from La Silla (weather allowing) 
since April 2008.
The spectral energy distribution (SED) obtained with GROND between
400-2400 nm allows us to not only find high-$z$ candidates 
\citep{gkf09, ksg11}, but also measure the extinction and the power law slope
\citep{gkk11}.
In the majority of all cases, this allows for a relatively 
accurate extrapolation of the SED into the sub-mm band, and
consequently a prediction of the flux for sub-mm instruments,
provided that $\nu_m$ has already passed the sub-mm band (which will
be shown below to be the case for the majority of GRBs after about 1 day).

\begin{figure}[th]
\includegraphics[width=9.3cm]{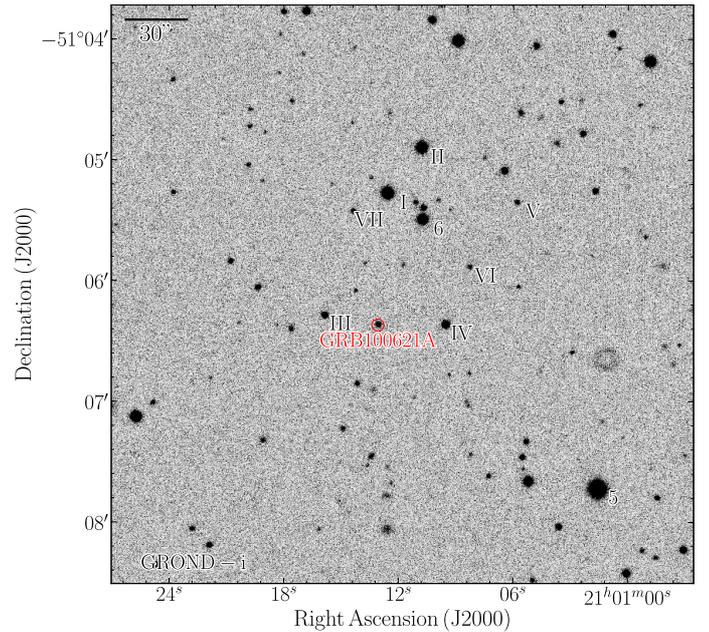}
\caption{GROND $i'$-band finding chart of GRB 100621A, including the
 photometric comparison stars (roman and arabic letters). 
North is up, and East to the left.
\label{grondfc}
}
\end{figure}

GRB 100621A triggered the Burst Alert Telescope (BAT) on the \textit{Swift} satellite 
\citep{gcg04} on June 21, 2010 at $T_{\rm o}$ = 03:03:32 UT
\citep{ubb10}. The prompt emission consists of a bright (25000 cts/s peak
count rate in the 15-350 keV band), smooth, triple-peak 
burst with a duration of nearly 70 s.
\textit{Swift} slewed immediately and started taking data with the XRT and UVOT
telescopes at 76 s after the trigger. A bright X-ray afterglow was found 
at RA (2000.0) = 21\h 01\m 13\fss24, Decl. (2000.0) = -51\degr 06\amin 21\farcs7
with an error radius of 1\farcs7 \citep{ego10}. In fact, 
GRB 100621A has had the brightest X-ray afterglow ever detected:
with an initial count rate in excess of $\approx$140\,000 cts/s, it
saturated the XRT CCD for several minutes.
Starting  80 seconds after the burst, the X-ray light curve in the 0.3--10 keV
band can be modelled with 4 power-laws\footnote{Throughout this paper, 
we use the definition F$_{\nu} \propto t^{-\alpha} \nu^{-\beta}$ where 
$\alpha$ is the temporal decay index, and $\beta$ is the spectral slope.},  
with decay indices and temporal breaks as follows: 
$\alpha_1$  = 3.87$\pm$0.02, $t_{\rm break1}$ = 439$\pm$10 s,
$\alpha_2$  = 0.51$^{+0.02}_{-0.03}$, $t_{\rm break2}$ = 6.2$^{+1.2}_{-0.5}$ ks,
$\alpha_3$  = 1.0$\pm$0.1, $t_{\rm break3}$ = 122$^{+0.13}_{-0.21}$ ks,
and $\alpha_4$  = 1.73$\pm$0.08 \citep{use10}.

GRB 100621A was also detected with INTEGRAL/SPI-ACS\footnote{http://www2011.mpe.mpg.de/gamma/instruments/integral/spi/acs/ grb/trigger/2010-06-21T03-03-26/index.html} 
and Konus-Wind, providing a time-integrated
spectrum with best-fit low-energy power law index $-1.7$, high-energy
index $-2.45$ and a peak energy $E_p$=95$^{+18}_{-13}$ keV \citep{gaf10}.
At $z=0.54$ and standard cosmology 
($H_{\rm o}$=70 km/s/Mpc, $\Omega_{\rm M}$=0.27,
$\Omega_{\rm \Lambda}$=0.73), this implies an isotropic energy release
of $E_{\rm iso}$ = (2.8$\pm$0.3) $\times$ 10$^{52}$ erg \citep{gaf10}.

Initially, no UVOT counterpart was detected,
and also rapid ground-based imaging with robotic telescopes
(like ROTSE, \cite{prg10}) did not find an afterglow. Prompted by the 
discovery of a very red afterglow with GROND \citep[][but see below]{unn10},
a spectrum taken with X-Shooter at the VLT determined a redshift of
z=0.542 \citep{mgt10}, and also faint UVOT detections were 
recovered \citep{use10}.

Here, we describe our multi-wavelength observations and results for 
GRB 100621A, and present an analyses of the data in the framework of the
fireball scenario.

\section{Observations}

\begin{table*}[th]
   \caption{Secondary standards used for the GROND data}
   \vspace{-0.2cm}
      \begin{tabular}{cccccccc}
      \hline
      \noalign{\smallskip}
    Filter &  Star I   & Star II      & Star III      & Star IV & Star V & Star VI & Star VII \\
           & 21 01 12.58 & 21 01 10.81 & 21 01 15.88 & 21 01 09.54 & 21 01 05.82 & 21 01 08.30 &21 01 14.38 \\
           & $-$51 05 17.2 & $-$51 04 54.6 & $-$51 06 17.4 & $-$51 06 22.2 & $-$51 05 21.5 & $-$51 05 53.6 & $-$51 05 25.8 \\
 \noalign{\smallskip}
      \hline
      \noalign{\smallskip}
  $g'$ & 16.60$\pm$0.05 & 16.28$\pm$0.05 & 18.54$\pm$0.05 & 20.14$\pm$0.05 & 20.34$\pm$0.06 & 20.38$\pm$0.06 & 19.49$\pm$0.05 \\
  $r'$ & 15.56$\pm$0.04  &15.64$\pm$0.04 & 18.09$\pm$0.04 & 18.58$\pm$0.04 & 19.44$\pm$0.05 & 19.70$\pm$0.04 & 19.15$\pm$0.04\\
  $i'$ & 15.29$\pm$0.04 & 15.48$\pm$0.04 & 18.00$\pm$0.04 & 17.33$\pm$0.04 & 19.18$\pm$0.04 & 19.55$\pm$0.05 & 19.10$\pm$0.04\\
  $z'$ & 15.05$\pm$0.04 & 15.31$\pm$0.04 & 17.93$\pm$0.04 & 16.69$\pm$0.04 & 18.90$\pm$0.04 & 19.39$\pm$0.04 & 19.00$\pm$0.04\\
  $J$  & 14.91$\pm$0.05 & 15.34$\pm$0.05 & 18.02$\pm$0.05 & 16.23$\pm$0.05 & 18.83$\pm$0.05 & 19.42$\pm$0.05 & 19.13$\pm$0.05\\
  $H$  & 14.76$\pm$0.06 & 15.31$\pm$0.06&  18.14$\pm$0.07 & 16.04$\pm$0.06 & 18.65$\pm$0.08 & 19.42$\pm$0.09 & 19.29$\pm$0.08\\
    \noalign{\smallskip}
     \hline
     \hline
     \noalign{\smallskip}
  Filter  &  Star 1=I   & Star 2      & Star 3      & Star 4=IV & Star 5 & Star 6 &  \\
       & 21 01 12.58 & 21 01 34.92 & 21 01 03.38 & 21 01 09.53  & 21 01 01.58 &  21 01 10.74  & \\
       & $-$51 05 17.2 & $-$51 05 59.3 & $-$51 03 26.6 & $-$51 06 22.5  & $-$51 07 43.8 &  $-$51 05 30.2  & \\
     \noalign{\smallskip}
      \hline
      \noalign{\smallskip}
   $K$ & 15.12$\pm$0.07 & 12.93$\pm$0.07 & 14.72$\pm$0.07 & 16.28$\pm$0.09 & 13.57$\pm$0.07 & 16.26$\pm$0.08 & \\
     \noalign{\smallskip}
      \hline
   \end{tabular}
   \label{compstar}
\end{table*}

\begin{table*}[th]
   \caption{APEX/LABOCA observations at 345 GHz in photometry mode}
   \vspace{-0.2cm}
      \begin{tabular}{ccccrc}
      \hline
      \noalign{\smallskip}
      Date & Time after GRB & On+Off time & Avg. $\tau$ &  Flux & Eff NEFD \\ 
       (UT)&     (days)     & (s)       &             & (mJy )& (mJy sqrt(s)) \\
      \noalign{\smallskip}
      \hline
      \noalign{\smallskip}
     Jun 22 04:38-05:30 & 1.0835 &  607 &   0.234    & 35.5$\pm$3.3 &   61.8 \\ 
     Jun 23 07:27-08:15 & 2.1996 &  600 &   0.358    & 23.6$\pm$3.8 &   64.0 \\ 
     Jun 25 07:51-08:42 & 4.2184 &  592 &   0.376    &  5.2$\pm$3.4 &   54.4 \\ 
     \noalign{\smallskip}
      \hline
   \end{tabular}
   \label{APEX}
\end{table*}

\begin{table*}[th]
   \caption{ATCA observations}
   \vspace{-0.2cm}
      \begin{tabular}{cccc}
      \hline
      \noalign{\smallskip}
      Date & Time after GRB &  Flux\,@\,5.5 GHz & Flux\,@\,9.0 GHz  \\ 
      (UT)  &     (days)     & ($\mu$Jy)         &   ($\mu$Jy)     \\
      \noalign{\smallskip}
      \hline
      \noalign{\smallskip}
     Jun 24 19:00 -- Jun 25 15:30 &  4.0910 & 137$\pm$17 & 150$\pm$28  \\ 
     Jun 25 15:30 -- Jun 26 12:00 &  4.9451 & 129$\pm$24 & 127$\pm$45  \\ 
     Jul 17 08:00 -- Jul 18 14:00 & 26.2083 & $-$43$\pm$85 & 49$\pm$100  \\ 
     \noalign{\smallskip}
      \hline
   \end{tabular}
   \label{ATCA}
\end{table*}

\subsection{GROND observations}

\begin{figure}[t]
\includegraphics[width=9.0cm]{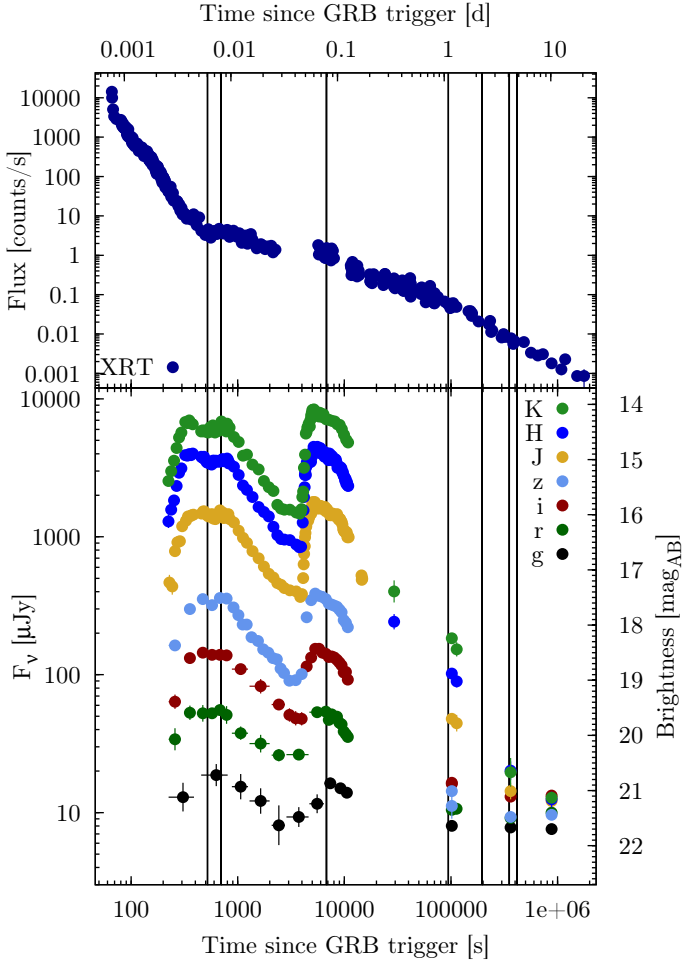}
\caption{Afterglow light curve of GRB 100621A as observed with \textit{Swift} in 
X-rays (top)
and GROND in its seven filter bands (bottom). The $J$-band data points 
at 14 ks are from SOFI imaging, and the $HK_{s}$-band data at 20 ks
from a GROND-observation in morning twilight at which the $J$-band
was already saturated by the rising Sun.
The 7 vertical lines 
mark the times at which spectral energy distributions have been 
extracted (see text and Fig. \ref{SEDs}).
\label{grondlc}
}
\end{figure}

Some of the GROND data of this burst, in particular the $J$-band light 
curve and the host measurements, have already been reported in \citep{kgs11}.
Here, we report the full data set, including the multi-band light curve,
and the SED evolution.

GROND exposures automatically started 230 s after the \textit{Swift} trigger, 
one of the fastest reactions of GROND@2.2\,m so far.
Simultaneous imaging in $g'r'i'z'JHK_{\rm s}$ continued for 3.05 hrs, and 
was resumed on nights 2, 4, and 10 after the burst. 
GROND data have been reduced in the standard manner 
using pyraf/IRAF \citep{Tody1993, kkg08b}.
The optical/NIR imaging was calibrated against the primary 
SDSS\footnote{http://www.sdss.org} 
standard star network, or cataloged magnitudes of field stars from the 
SDSS in the case of $g^\prime r^\prime i^\prime z^\prime$ observations or the 
2MASS catalog for $JHK_S$ imaging. This results in typical absolute accuracies 
of $\pm$0.03~mag in $g^\prime r^\prime i^\prime z^\prime$ and $\pm$0.05~mag in 
$JHK_S$. The light curve of the GRB 100621A afterglow in all 7 GROND filters is 
shown in Fig. \ref{grondlc}.

\subsection{Swift XRT data}

\textit{Swift}/XRT data have been reduced using the XRT pipeline provided
by the \textit{Swift} team.
The X-ray spectra were flux-normalized to the epoch 
corresponding to the GROND observations using the XRT light curves from 
\citet{ebp07, ebp09}. 
We then combined XRT and Galactic foreground extinction ($E(B-V)= 0.03$ mag; 
\cite{sfd98})
corrected GROND data to establish broad-band spectral energy 
distributions (SEDs)  which are shown in  Fig. \ref{SEDs}.

\subsection{NTT observations}

NTT/SOFI at La Silla was used to obtain NIR-spectroscopy. 
After recognizing the sharp drop in intensity at about $T_o$ + 10 ks
we took four 60-s  $J$-band images starting at 07:05 UT, on 21 Jun 2010.
While the results of the spectroscopy
are deferred to a later publication (these are of no relevance for
the purpose of this paper), the imaging provides an
additional photometric data point at a time when no GROND observations
were possible anymore due to visitor mode regulations. 
The SOFI images were reduced in the same manner as
the GROND $JHK$ data (actually within the same GROND pipeline),
and calibrated against the 2MASS catalog.

\subsection{APEX observations}

Since the SED slope, even after extinction correction, was rather steep,
the predicted sub-mm flux density of $\approx$50 mJy at 1 day after the GRB
led us to submit a DDT proposal to ESO for observations
with LABOCA \citep{siringo09}  on the Atacama Pathfinder 
Experiment APEX\footnote{APEX is a collaboration between the 
Max-Planck-Institut f\"ur Radioastronomie, the European Southern Observatory 
and the Onsala Space Observatory.} which was accepted at 
very short turn-around time.

LABOCA, the ``Large APEX Bolometer Camera'',  is an array of 295 
composite bolometers. The system is optimized to work at the central
frequency of 345 GHz with a bandwidth of about 60 GHz. 

The first APEX/LABOCA observation was obtained 1.08 days after the GRB, 
leading to a clear detection. 
Two other additional observations were performed at 2 days (another clear
detection) and 4 days (upper limit only) after the GRB.
This makes GRB 100621A one of the rare cases with a sub-mm light curve
(see section 5.3).
All these observations were carried out in photometry mode.

Immediately after the first epoch observation (done in photometry mode), 
we obtained at 5:32-6:26 UT a complementary observation of GRB 100621A in
mapping mode, for an exposure of 7x 420 s and reaching a 
1$\sigma$ sensitivity of 14 mJy/beam. While no source
was detected in this less sensitive observing mode, it verifies that there is
no strong, unrelated source close to the GRB position,
which otherwise could cause problems with the photometry mode data.

Reduction of the photometric data was done with the software BoA
\citep{schu12}  using standard 
routines for photometry mode. Subscans were checked individually before
averaging them together
in order to identify and remove outliers. The raster map was reduced
with the CRUSH \citep{kov08}
software package. Flux density calibration was done against Neptune,
G45.1 and B13134.

\subsection{ATCA observations}

In response to the initial detections of a bright afterglow of GRB 100621A 
\citep{ubb10, ego10, unn10, mgt10},
we also initiated observations of GRB 100621A 
with the Australia Telescope Compact Array (ATCA) in Narrabri, Australia, at 
the frequencies of 5.5 and 9.0 GHz with an observing bandwidth of 2 GHz. The
observation sessions were carried out between 24-26 June and 17-18 July 2010.
The radio counterpart of the afterglow of GRB 100621A was detected during 
the sessions carried  
out in June 2010 at both 5.5 and 9.0 GHz at a position coincident with those 
of the X-ray and optical counterparts, and it was undetected in the July 2010 
session. 

It is possible that the observed decay
between the first and second epoch, or part thereof, 
is due to interstellar scintillation, rather than due to the intrinsic
decay of the afterglow. Otherwise, the fading at 5.5 GHz would have been
rather early, indicating a rather low energy and/or $\epsilon_B$.

\section{Overall light curve behaviour}

The overall temporal evolution of the afterglow at X-rays and the
optical/NIR is shown in Fig. \ref{grondlc}.
The light curve in the X-ray band is very typical of X-ray afterglows
as seen by \textit{Swift}, with a steep decline (slope of --3...--4)
during the first $\approx$400 s, followed by a shallow decay until 
about 122 ks, after which the decay steepens to a slope of 1.73$\pm$0.08
\citep{use10}. In contrast, the temporal evolution of the optical/NIR 
afterglow is considerably more complex.
From the start of the GROND exposures at 230 s post-trigger, the
light curve shows a rapid rise with $\alpha_1 = -4.3^{+1.0}_{-0.6}$.
From about 400 s (consistent within errors with the end of the
steep X-ray decline) to about 700 s, the light curve is more or less flat
($\alpha_2 = 0.05\pm0.05$) with just a few wiggles. The sub-sequent decay has 
$\alpha_3 = 1.15\pm0.15$,
significantly steeper than the X-ray decay at
that time. After a short flattening (3--4 ks post-trigger), 
an extremely steep increase in optical/NIR brightness is observed 
from 4 to 5 ks after 
the trigger which has also been reported by the SIRIUS/IRSF team \citep{nss10}.
This intensity jump is larger in the NIR than in the optical, 
reaching an amplitude of 1.9 mag in the $K_{s}$-band. A formal fit results
in $\alpha_4 = -14^{+1.3}_{-0.6}$, the steepest flux rise we have ever
seen in a GRB afterglow (at any time), both in the literature as well as
in our GROND data over the last years. After a short-lived (5--9 ks)
slow decline with $\alpha_5 = 0.42\pm0.05$, a steep decay with
$\alpha_6 = 2.3\pm0.1$ sets in which flattens into the host flux level
at around 3$\times$10$^5$ s.

\section{Broad-band afterglow SED modelling}

\subsection{Fitting framework, spectral breaks and cooling stage}

In the following, we will analyse our data in the framework of the
fireball scenario, in particular in the formalism as described in 
\cite{grs02}. From the single-epoch spectra in certain wavebands
we can derive some basic boundary conditions, as follows.

\begin{figure*}[th]
\centering{
\includegraphics[width=12.2cm,angle=90]{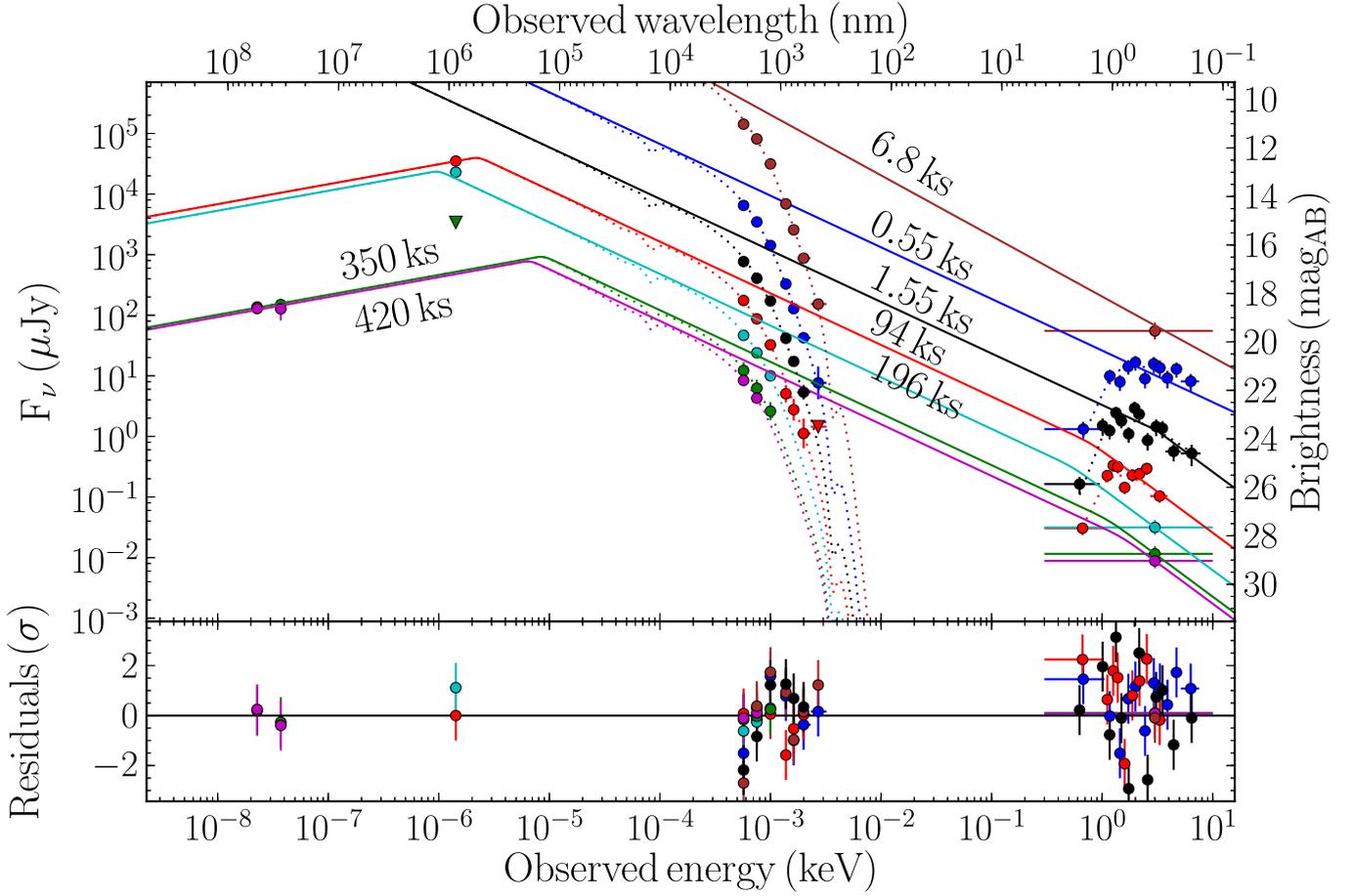}}
\caption{Multi-epoch SEDs (different colours) of the late-time afterglow of
GRB 100621A as measured by
 \textit{Swift}/XRT (right; not $N_{\rm H}$-corrected), 
GROND (middle; not $A_{\rm V}$-corrected)), 
APEX/LABOCA (middle left)  and ATCA (far left),
together with a broad-band model which fits all data available for 
the given epoch.
The times of these SEDs are marked with vertical lines in Fig. \ref{grondlc},
and the resulting break energies given in Tab. \ref{breaks}.
Since the optical/NIR and X-ray fluxes in epochs 1--3 are
very similar, epoch 3 (jump component at 6.8 ks) has been
scaled upwards by a factor of 20, and epoch 2 (flares) 
down-scaled by a factor of 4.
The curvature in the GROND data is due to 
strong extinction of the afterglow light in the host galaxy (dotted line).
The breaks seemingly show erratic variations in frequency -- see text for an
interpretation. Note in particular, that we consider the fits in this
plot not to be the final physical interpretation of the data, as it
links emission components at different wavelength regions which we argue
in the text to not belong together.
\label{SEDs}
}
\end{figure*}

We start by fitting the GROND-data of the first 1 ks on its own. 
The SED built from
the 7 GROND filters is very steep, but also clearly curved (right of center 
in Fig. \ref{SEDs}),
indicating substantial host-intrinsic extinction. As is standard practice,
we apply a power law (as one segment of the fireball scenario) and fit 
the power law slope together with the dust extinction $A_{\rm V}$ in the 
rest-frame of the GRB (z=0.542).
The resulting best-fit spectral slope in the optical/NIR range 
(well before the strong intensity jump at 4 ks)
is measured to be $\beta \sim 0.8\pm0.1$.
Any slope flatter than $\beta \sim 0.7$, in particular the theoretical
prediction of  $\beta$ = 0.5 for certain conditions \citep{grs02}, is safely
excluded by the data (note that there is no ambiguity with the
intrinsic host extinction $A_{\rm V}$ = 3.6 mag, see next section).

Similarly, we fit the \textit{Swift}/XRT data on its own, and
reproduce a slope of $\beta_{\rm X} = 1.4\pm0.2$ and 
$N_{\rm H} = 6.5\times 10^{21}$ cm$^{-2}$ as given in \cite{use10}.
Since we observe a steeper spectral slope in X-rays, this excludes
the fast cooling options (``spectrum 4 \& 5'' in \cite{grs02}) at early times,
and by construction (evolution from fast cooling to slow cooling) 
also at late times.

Since the steepest possible fit to the GROND optical/NIR data is 
$\beta \sim 1.1$, but the X-ray spectrum 
is significantly steeper than this, we are forced to introduce a break
between the optical/NIR and X-ray data at intermediate times.
Since at early times a single power law for the combined GROND and 
\textit{Swift}/XRT data is sufficient, this break 
has moved into the covered bandpass. 
We interprete this break as $\nu_c$, as the observed slope difference
of 0.6$\pm$0.2 is consistent with the predicted value of 0.5.
If this break had moved from the
infrared through the optical, the optical/NIR slope should have
gotten bluer -- which is not observed. In addition, the X-ray spectrum
steepens, consistent with $\nu_c$ moving from high energies down through
the X-ray band. We therefore conclude
that the external density profile is constant (ISM-like).

The simultaneous 5.5 and 9.0 GHz measurements at 4 and 5  days after the GRB
suggest a relatively flat slope of $\beta  \approx -0.25$ (with relatively 
larger error), 
implying that the self-absorption frequency $\nu_{sa}$ is below 5.5 GHz.
Again, as we observe (at certain times) a spectral break between
the optical/NIR and the X-ray bands, an interpretation according to 
``spectrum 2'' or ``spectrum 3''
 \citep{grs02} with the self-absorption frequency slightly above 9.0 GHz 
(i.e. near its peak at the transition between $\nu^{(1-p)/2}$/$\nu^{-p/2}$ to 
$\nu^{5/2}$) is impossible, as there would be no further break at higher 
frequencies.

Thus, we are left with the option of ``spectrum 1'' \citep{grs02}, for which
the fireball prediction is $\beta$ = $-$1/3 above
the self-absorption frequency, in reasonable agreement with the measured 
$\beta$ = --0.25. While this conclusion is formally valid for the
time of the radio measurements at 4 and 5 days after the GRB, 
any other spectral phases (``spectrum 2'' to ``spectrum 5'' from \cite{grs02})
have been excluded by the above considerations.
We therefore conclude that already at early times ($T_o + 500$ s)
the afterglow is in the slow cooling phase.

We therefore continue with the conceptual interpretation of slow cooling
throughout our full data set, and the frequency ordering as
$\nu_{sa} < \nu_m < \nu_c$, i.e. with the break between the optical/NIR and 
the X-ray part of the spectrum interpreted as
the cooling break $\nu_{c}$, and the break long-wards of the
optical/NIR as the injection frequency $\nu_{m}$.

We will model the SED at various epochs with a  three-component power law,
with slopes $\beta_1$ describing the radio range, $\beta_2$ the
GROND range, and $\beta_3$ the X-ray range. According to the
standard prescription \citep{grs02}, we fix the slope difference to 0.5 
around the cooling frequency  $\nu_{c}$, i.e.
$\beta_3$ =  $\beta_2$ + 0.5.
We also fix $\beta_1$ = -1/3, due to the otherwise large effect on $\nu_m$.
The three power law 
segments are smoothly connected with a fixed smoothness parameter of 15
\citep[see][]{bhr99}.

\subsection{Broad-band SED fitting}

For the following discussion, let us define 7 epochs which 
are sequentially in time:
epoch 1 = 450--600 s (diagonal-hatched region in Fig. \ref{flares})
epoch 2 = the sum of the time intervals 650--750 s, 900--1150, 
   1350--1800 s (cross-hatched regions in Fig. \ref{flares}),
epoch 3 = 5.5--8.5 ks,
epoch 4 = 94 ks,
epoch 5 = 196 ks,
epoch 6 = 352 ks,
epoch 7 = 416 ks,
where the last three epochs are primarily determined by the times of the
APEX and/or ATCA observations. In these latter three cases the optical 
flux has been determined by
interpolating the GROND light curve which looks pretty smooth at these 
late times. The last three GROND epochs come with 
considerable systematic uncertainty due to the host subtraction.
Due to the bright X-ray emission even at late times, no 
assumptions on the slope of the X-ray spectrum had to been made.

A fit of these seven SEDs with the assumptions as listed at the end
of the previous section and using all the available data at a given
epoch is shown in Fig. \ref{SEDs}. The most obvious result is that
the injection frequency (and there are good reasons why this 
is not a different break frequency, see above) moves to higher 
frequencies between epoch 5 (196 ks) and 6 (352 ks). This evolution
is inconsistent with any prediction of the fireball scenario.
While this is not a reason to condemn the fireball scenario,
we discuss two possible options to explain this behaviour,
both within the framework of the fireball scenario: \\
\noindent (1) 
If one relaxes the usual assumption that the microphysical parameters
are constant, the break frequencies would follow a more complicated
evolution than described in \cite{grs02}. While such a recourse has
been offered for the description of selected GRBs \citep[e.g.][]{fgs12},
in the present case one would have to invoke an increase of 
$\epsilon_{\rm e}$ proportional to $t^1$, or of
$\epsilon_{\rm B}$ as fast as $t^3$...$t^4$.
Moreover, this temporal evolution
would be required only for the time between epochs 5 and 6,
but not for the evolution as seen between epoch 4 to 5, or 6 to 7.
Thus, we consider this option physically implausible. \\
\noindent (2) Another option is that the true model, which results in 
the determination of the break 
frequencies, contains two (or more) different emission components which 
dominate at different frequency bands, or at different times. 
Already relatively small changes in flux of one component would
lead to substantial changes in the break frequencies, even at constant
slopes. A good example in our case is the epoch 3: assigning either
all observed X-ray flux or just 50\% of it (because the other 50\% 
might be the normal underlying afterglow) to the component which
produces the large intensity jump in the optical/NIR will change
the best-fit cooling break frequency by one order of magnitude.

\begin{table}[ht]
  \caption[]{Break energies as derived from the SED fitting as shown in 
  Fig. \ref{SEDs}; but see text.  \label{breaks}}
   \vspace{-0.2cm}
   \begin{tabular}{cccc}
   \hline
   \noalign{\smallskip}
   SED epoch & Time           & $\nu_m$ & $\nu_c$ \\
             &  (ks)         & (meV)   & (keV) \\
   \noalign{\smallskip}
   \hline
   \noalign{\smallskip}
    1 & 0.52   & $<$550            & $>$8 \\
    2 & 0.7$^*$& $<$550            & 2.9$^{+0.6}_{-0.5}$ \\
    3 & 6.8    & $<$550            & $>$8 \\
    4 & 94     & 2.9$^{+0.9}_{-0.5}$ & 0.003--0.1 \\
    5 & 196    & $<$2.0              & 0.003--0.1 \\
    6$^{**}$ & 352    &10.6$^{+3.1}_{-2.1}$ & 0.003--0.1 \\
    7$^{**}$ & 416    & 8.1$^{+2.5}_{-1.6}$ & 0.003--0.1 \\
   \noalign{\smallskip}
   \hline
    \noalign{\smallskip}
  \end{tabular}

$^*$ This is the center of the first of three intervals - see text.\\
$^{**}$ For these two epochs, our formal fit values for $\nu_m$
  are considered unphysical and thus likely an indication that the radio 
  and optical/NIR emission
  stem from different components -- see text. 
\end{table}

Thus, we conclude that a model-independent analysis  of our data set
is largely impossible, despite the broad frequency coverage and
the multiple epochs available in all frequency bands. Moreover,
as described above, the behaviour of the GRB 100621A afterglow is so 
complex that we are also not able to test some predictions of
(for example) the fireball scenario by our multi-epoch SEDs.

Instead, the only approach left is to develop an interpretation
as simple as possible within a given framework (and we chose the fireball
scenario for this) which describes the data to a large (possibly full)
extent. 
In what follows we use our data together with some basic arguments
derived from the fireball scenario to disentangle the complex
behaviour of the GRB 100621A afterglow into several different
components, the sum of which explain the observed features.
Our driving principle was to minimize the number of assumptions,
as well as emission components.
This is likely not a unique description, and a more sophisticated
interpretation is not excluded. 

We consider three different components: \\
(1) a canonical underlaying afterglow, \\
(2) flares during the first 1000 s, and \\
(3) a jump component, most prominently visible in the optical/NIR
  at 5.5-8.5 ks. \\
\noindent Each of these components is allowed to have a different 
electron distribution $p$,
and a different set of microphysical parameters such that the
break frequencies in each are different. For most of the time, at least two of 
these three components overlap, and care has to be taken to assess
which of the components dominates at which time or in which spectral range.
Our results, discussed below, suggest the following
superposition of components, where the break frequencies are given
for the dominant component in that frequency band: 

\begin{itemize}

\item epoch 1: optical/NIR and X-rays dominated by canonical afterglow,
  sub-mm and radio unconstrained; 
 neither $\nu_c$ nor  $\nu_m$ for SED of canonical afterglow are covered.

\item epoch 2: optical/NIR dominated by flares, X-rays are superposition
  with canonical afterglow, sub-mm and radio unconstrained;

\item epoch 3: optical/NIR dominated by jump component, X-rays
  are $\approx$50:50 superposition of canonical afterglow and jump component,
 sub-mm and radio unconstrained;
 neither $\nu_c$ nor  $\nu_m$ of jump component covered.

\item epoch 4/5:  optical/NIR dominated  by jump component,
X-rays dominated by canonical afterglow, sub-mm is likely the jump component;
$\nu_m$ of jump component in sub-mm.

\item epoch 6/7:  optical/NIR still dominated  by jump component,
X-rays and radio dominated by canonical afterglow, sub-mm not constrained;
$\nu_m$ of the SED of canonical afterglow is in the radio.

\end{itemize}

\subsubsection{Epochs 1 and 2 \label{sectep12}}

\begin{figure}[ht]
\hspace{-0.4cm}\includegraphics[angle=270,width=9.6cm]{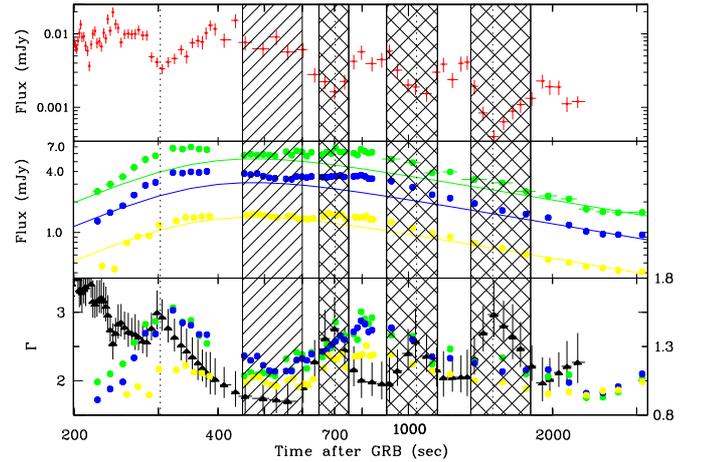}
\vspace{-0.2cm}
\caption{Comparison of the fluxed X-ray light curve at 10 keV (top panel), 
the GROND $J$ (yellow), $H$ (blue), $K_{s}$ (green)
bands (middle), and in the bottom panel the photon index of the X-ray 
spectrum (black, left $y$-axis scale) and the residuals of the model fit 
(see text)
to the GROND JHK data (color as in the middle panel, right y-axis scale).
The diagonal-hatched region denotes epoch 1, and the cross-hatched regions
epoch 2. The dashed vertical lines mark the maxima in the photon spectral
index $\Gamma$ ($\Gamma = \beta + 1$) to guide the eyes.
\label{flares}
}
\end{figure}

At first glance, the rise time in the optical is too fast for
a forward shock \citep{pav08}, and the temporal and spectral parameters are
not consistent with any closure relation (neither wind nor ISM
density structure, with either standard or a jetted afterglow).
Also, the subsequent part of the optical/NIR light curve
($T_o$+300 to $T_o$+600 s) is surprisingly flat. 
However, we note that the X-ray spectrum oscillates on a few hundred
seconds timescale between a steep ($\beta \sim 1.3$) and a
flat ($\beta \sim 0.8$) slope during the first few ks after the GRB. 
More interestingly, two of the three times
of steep spectral slopes coincide with flux depressions in the (fluxed) X-ray 
light curve, and flux enhancements (which could be described as optical flares)
 in the GROND data (lower panel in Fig. \ref{flares}, 
at 300 and 700-800 s).
This suggests that the evolution of the afterglow between
$T_o$+200 s to $T_o$+2000 s is the superposition of two components,
a ``normal'' afterglow and a flare component.

In order to disentangle these two components, we fit the 
X-ray spectral index evolution (lower panel of Fig. \ref{flares}) 
with a constant plus a number of separate Gaussians, 
whenever the spectral index deviates
more than 3$\sigma$ from the constant.
 We then apply a model composed of the rise and decline of a forward 
shock and the multiple Gaussians as derived in the previous step to the  GROND 
light curve, 
now with fixed times of occurrence of the Gaussian components, 
but allowing 
different width and normalizations. Due to better temporal resolution
and S/N-ratio we concentrate on the $JHK_{s}$ data. The residuals of such a fit 
without the Gaussians, 
i.e. the best-fit Gaussians to the GROND light curve on top of the forward shock
are overplotted over the X-ray slope variation
in the lower panel of  Fig. \ref{flares}. While there is no perfect
agreement in all slope-oscillations, there is a surprisingly tight coincidence
in the first two, at $T_o$+300 s and $T_o$+700 s.
The results of this exercise are:
\begin{itemize}
\item the early rise in the GROND light curve is likely dominated
by a flare, making 
the rising slope of the light curve
particularly steep; when including a flare
at $T_o$+300 s in the fit, 
the rise of the ``normal'' afterglow in the on-axis case
is consistent with $t^{2}$, suggestive of the canonical forward shock.
This is additional evidence for a constant density profile, as the rise
in a wind profile would be much slower ($t^{0.5}$ to $t^{1.0}$) \citep{pav08}.
\item The relatively flat light curve during the interval at 300--800 s
after the GRB trigger is due to the contribution, and likely superposition,
of flares. Once subtracted, the decay of the standard afterglow
is flatter, namely $\alpha = 0.69\pm0.03$, where a systematic error of
$\pm0.05$ should be added due to the ambiguity of choice of the strength 
and width of the flares. 
\item The optical/NIR emission during the intervals 
$T_o$+450 s -- $T_o$+600 s (and $T_o$+2500 s -- $T_o$+3000 s)
are the only times when GROND sees ``normal'' afterglow at early times.
This corresponds to our definition of epoch 1. A combined fit of the
GROND and \textit{Swift}/XRT data results in a single power law of 
$\beta = 0.81\pm0.02$ with no spectral break being preferred over a 
fit with a break. Taking the corresponding Galactic contributions
into account, the best-fit rest-frame dust extinction and
effective hydrogen  absorption are 
$A_{\rm V} = 3.65\pm0.06$ mag, and 
$N_{\rm H} = (1.8\pm0.3) \times 10^{22}$ cm$^{-2}$.
The inferred slope above $\nu_c$ would be $\beta_3 = 1.31$, with
 $\nu_c > 8$ keV, and the corresponding electron spectral index
$p = 2.62\pm0.04$.
\item The peak of the forward shock is at 380$\pm$30 s, corresponding
to an initial Lorentz factor of 71$\pm$3
(according to the new prescription of \citet{gng12} which returns
values about a factor two lower than the previously used ones like
\citep{mol07}).
\item The emission during the flares is much steeper in X-rays,
with best-fit spectral slopes in the 1.2--1.8 range. A combined
GROND and \textit{Swift}/XRT fit results in the need of a spectral break
(at $\sim$3 keV), with low- and high-energy power law slopes of
$\beta_2 = 0.86\pm0.06$ and  $\beta_3 = 1.36\pm0.06$ 
(with fixed $\Delta\beta = 0.5$).
It is interesting to note that the spectrum alternates four times
during the first 1000 s between this steep flare spectrum and the
flatter ``normal'' spectrum.
\end{itemize}

Considering these results for the ``normal'' afterglow, i.e. with
$\alpha_O = 0.69\pm0.06$,
$\alpha_X = 0.74\pm0.02$ (note that we deviate from \cite{use10} in that
we fit the $T_o$+700 s to $T_o$+100 ks interval with one straight 
power law, but omit the higher-flux portion at $T_o$+6 ks, see below 
and Fig. \ref{xlc}),
and $\beta_2 = \beta_3 = 0.81\pm0.02$ with inferred $p=2.62$,
we find consistency in the optical/NIR and X-ray decay slopes, 
but also note that this is much flatter than one
would expect with the canonical closure relations for a standard
afterglow with the given $p$ in either wind ($\alpha=1.72$) or 
ISM ($\alpha=1.22$) environment.
This suggests some form of energy injection. If the addition of 
energy is a power law in the observer time, $E_i(<t) \propto t^e$,
then the flattening is by $\Delta\alpha = e*1.41(0.91)$ for a ISM (wind)
density profile at $\nu < \nu_c$ \citep{pmg06}. Thus, 
with $e = 0.35-1$, depending
on the circumburst density structure, consistency could be reached.
As we will show below, our data are not compatible with a wind medium,
so we adopt an energy injection according to $E_i(<t) \propto t^{0.35}$
until $T_o$ + 4 ks.

\begin{figure}[th]
\includegraphics[angle=270,width=9.4cm]{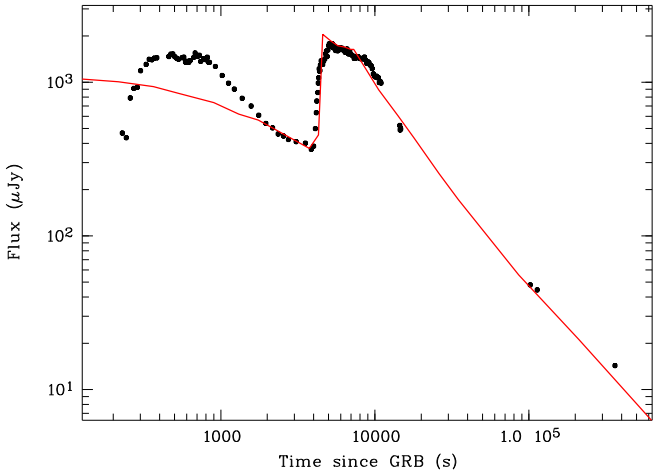}
\caption{Early part of the GROND J-band light curve with a (slightly
stretched in time) model of 
the two-shell collisions overplotted (case 4, Fig. 7 in \cite{vlasis11}).  
Though this model was not aimed at reproducing the behaviour of the
GRB 100621A afterglow, the similarity of  the rise, structure at the peak 
and the decay slope is striking. The early part of the model should be
ignored, as it depends on the relative timing of the forward shock
of the first shell, the ISM density and initial Lorentz factor.
\label{lcvlasis}
}
\end{figure}

\subsubsection{Epoch 3 - the intensity jump \label{sectep3}}

While the short interval of the steep rise between 4.0-4.5 ks after the trigger 
is not covered by the \textit{Swift}/XRT due to Earth limb constraints, the time of the
first optical peak including the following slow decay
phase until T$_o$+8 ks is covered with \textit{Swift}/XRT observations, 
but shows only a marginal X-ray flux enhancement, on the 
order of 50\% relative to earlier and later times. 
This is in full agreement with the chromaticity seen within
the GROND band (after host subtraction and extinction correction), 
where the flux enhancement ranges between 200\% (0.8 mag) in the $g'$-band 
and 570\% (1.9 mag) in the $K_{s}$ band, implying a very red/soft
spectral shape. A combined GROND/XRT spectral fit of the overlapping
time interval 5.5--8.5 ks returns a single power law as best fit with 
a slope of $\beta =0.98\pm0.02$ when fitting all X-ray flux, or
 $\beta =1.0\pm0.03$ 
when fitting just 50\% of the  X-ray flux (under the 
assumption that the other 50\% belongs to the ``normal'' afterglow).
Two notes are in order: 
First, the SED can also be fit with a broken power law, with the break
somewhere between the GROND and the \textit{Swift}/XRT data. However, the improvement
in reduced $\chi^2$ is only marginal, so we adopt the simpler model.
Consequently, we assume $\nu_c > 8$ keV in the following.
Second, the above decomposition assumed similar spectral slopes, which
cannot be proven unambiguously. However, if the X-ray spectrum of the jump 
component would have been steeper by 0.5, with correspondingly $\nu_c$ 
being between the GROND and \textit{Swift}/XRT ranges, then one would
not have expected to see any X-ray flux increase at all.

\begin{figure}[th]
\includegraphics[angle=270,width=9.8cm]{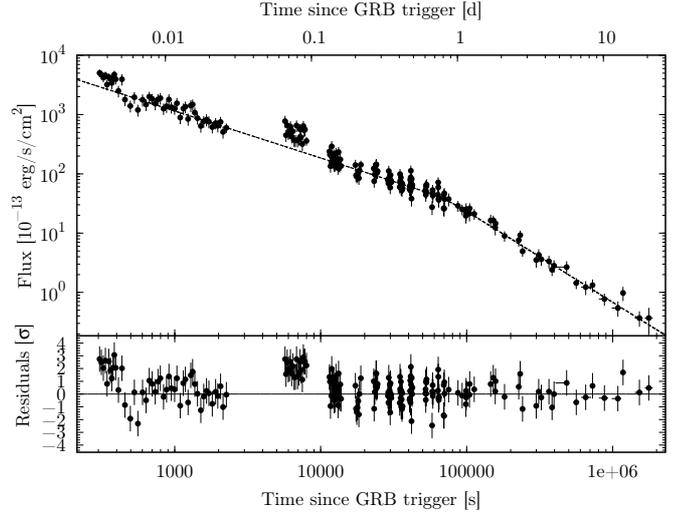}
\caption{X-ray light curve of the GRB 100621A afterglow with a 
broken power law fit, ignoring the enhanced emission at 5--8 ks
which we assign to the jump component (see subsection \ref{sectep3}).
The decay slope from around 1 ks nicely continues until 80 ks,
when it steepens to $\alpha = 1.54\pm0.06$.
\label{xlc}
}
\end{figure}

The overall shape of the rise, short shallow decay and subsequent fast
decay is very similar to the behaviour of the afterglow of GRB 081029 
\citep{ngk11},
where an analogous behavior has been associated with the 
intrinsic properties of the GRB and not to changes in the intervening dust 
content. In the meantime, but independent of these observations,
\cite{vlasis11} have presented numerical simulations of the collision of
an ultra-relativistic shell in a constant density environment with the 
external forward shock,  which
produce similar flare light curves: Fig. \ref{lcvlasis}
shows their case 4 model (with 2\degr\, half-opening angle; from their Fig. 7)
overplotted over
the GROND J-band light curve. In this scenario, the fast rise occurs
when a second shell reaches the back of the first, self-similar Blandford-McKee
shell. The steepness and amplitude of the rise depend on the half-opening angle 
of the jet, the Lorentz factor of the two colliding shells, and likely
more parameters like the energy, the occurrence time relative to the
jet break, and $\epsilon_B$. A parameter study much more extensive than
that in \cite{vlasis11} is needed in order to be able to derive some of 
these parameters (or ranges thereof) for GRB 100621A. However,
 a qualitative conclusion would likely be that GRB 100621A has
a large Lorentz factor or a small half-opening angle, or both.
Among the sample of a handful of GRBs showing such features \citep{grei11}, 
GRB 100621A shows the steepest rise in time: a formal fit with $T_o$
at the GRB trigger results in $\alpha_{\rm rise}$ = 14
(which due to its late appearance is also insensitive on any possible
change in $T_o$ of the fit)!

According tto  \cite{vlasis11},
the rather flat part after the jump is then due to the
merging of the two shells, the heating of which compensates the
fading flux from the forward shock of the first shell. 
After the jump, the light curves should 
follow the predicted slopes for the normal, single forward shock,
but at a higher intensity level due to the additional energy injection
by the colliding second shell.
While this is difficult to convincingly test with our data since 
the normal decay is not accurately enough constrained,
the rise and the
observed structure in the flat part of the light curve is surprisingly
similar to the modelling in \cite{vlasis11}, particularly their Fig. 6.
We defer a more detailed comparison of this behaviour in GRB 100621A 
with this shell-collision model to a future paper. 

\subsubsection{The light curve beyond 20 ks}

We have shown in sub-section \ref{sectep12} that the normal afterglow
decay slope in the optical/NIR at a few ks after the GRB was 
$\alpha_O = 0.69\pm0.06$. An extrapolation of this decay at the same
decay rate, i.e. with continued energy injection at the same temporal 
rate, underpredicts the later GROND data by at least a factor of 2. 
Thus, the rate of energy injection would have had to increase over the 
early rate, if it were to explain the optical/NIR emission at 
$T_o$+20 ks. We consider this unlikely, and thus conclude that
at late times, i.e. $t > T_o$+20 ks, the optical/NIR fluxes are
dominated by the process which led to the huge intensity jump at 4 ks.
As the spectral shape of this emission was redder than that of the 
``canonical'' afterglow, this statement will also be true for the sub-mm 
and radio bands (see next sub-section). 
At X-rays, we have shown in the previous sub-section 
that the contribution of the large intensity jump was marginal, at 
most 50\%, during the peak emission of the intensity jump.
If the X-ray emission associated to the jump component subsequently dropped
the same way as the optical emission, 
then it faded by a factor of 20 in the
interval from $T_o$+10 ks to $T_o$+30 ks. The total X-ray emission
faded by just a factor of 2, implying that the X-ray emission
beyond about $T_o$+10 ks can be solely attributed to the normal 
afterglow.

The fit to the X-ray light curve, using a broken power law and 
ignoring the enhanced emission at 5--8 ks, describes the overall
behaviour very well. The break time is derived to be 80 ks, 
at which point the decay steepens to $\alpha = 1.54\pm0.06$ (Fig. \ref{xlc}).

This steepening of the light curve could be due to the cessation of
the energy injection. However, for our value of $\beta$, 
a full cessation should lead to the
canonical decay slopes of $\alpha=1.72$ (wind) or $\alpha=1.22$ (ISM)
in the standard afterglow scenario, or $\alpha=1.96$ or steeper for
any jet model (see below).
Thus, only a partial cessation of energy injection 
would be a viable solution.

Alternatively, it could be the passage of the cooling break at continued
energy injection.
This would not work for the standard afterglow scenario
of a spherical afterglow (i.e. $\Gamma > 1/\theta$, where $\theta$ is the
jet half-opening angle), since the predicted slope change is just
$\Delta\alpha = 0.25$. However, the predicted change is larger for
a jetted outflow. Following \cite{pmg06}, we consider two options:
(i) a jet whose edge is visible and which does not expand laterally,
and (ii) a jet with sharp edges which spreads laterally and is observed
when $\Gamma \times \theta < 1$. In their eqs. 34 and 35, \cite{pmg06}
provide the flattening of light curves due to energy injection 
for the frequency range above and below $\nu_c$. 
For option (i), the slopes depend on the circumburst medium density profile.
Thus, we have three cases, each with a separate closure relation
above and below $\nu_c$. We start with the three cases for $\nu < \nu_c$
and determine $e$, the power of the energy injection (see above) such
that the observed early decay slope of $\alpha = 0.72$ is reproduced
(we choose to take the value consistent with both our measured 
$\alpha_O$ and $\alpha_X$, though this would not change our upcoming
conclusion). With each of the three different values of $e$, we then
check the predicted slope at $\nu > \nu_c$ for each of the three cases.
Option (i) in the constant density environment returns the steepest
slope, with $\alpha_{\rm pred} = 1.2$ (for $e=0.75$). This is still flatter
than the observed $\alpha_X = 1.54\pm0.06$ (Fig. \ref{xlc}). 
The predicted slope depends only very weakly on $\beta$, so also the
trend of steepening $\beta_X$ towards the end of the observed X-ray
light curve will not lead to consistency. 
We note that in this interpretation the energy injection is still
active at the end of the X-ray light curve, i.e. at $2\times 10^6$ s,
as we see no further steepening to a slope of $\alpha > 2.2$ (depending
on any further softening of $\beta_X$).

Last, but not least, we note the coincidence of the measured slope
of $\alpha$ = 1.54$\pm$0.06 and the predicted $\alpha_X$ = 1.48
for the decay of the $\nu > \nu_c$ part of the afterglow in the spherical
case. Thus, the steepening of the X-ray light curve at 80 ks
could be due to the combination of both, cessation of energy injection
AND passage of the cooling break in an ISM environment for an afterglow
which is still in its spherical expansion phase (i.e. $\Gamma > 1/\theta$)
when the collimation is not yet detectable. 
Admittedly, the need for this coincidence is not an attractive solution.
 At the moment, we have
no more satisfactory explanation for the amount of the steepening of
X-ray light curve at $T_o$+80 ks.
However, such break in the X-ray light curve is very common in the
sample of 
$\approx$700 \textit{Swift} GRB afterglows, 
and thus a more generic problem \citep{nkg06} rather than
related to the specifics of GRB 100621A.

\subsubsection{Epochs 4 and 5}

The two APEX/LABOCA detections correspond to a flux decay according to
$\sim t^{-0.5}$ which then must accelerate considerably
in order to be compatible with the upper limit at epoch 6. 
In the standard fireball scenario, the maximum in the sub-mm light curve 
is associated with the
passage of the injection frequency. Since the observed decay slope
is still considerably flatter than the expected $t^{3(1-p)/4}$
for  $\nu < \nu_m$, the injection frequency of the
dominating component should be near the LABOCA observing frequency 
during epochs 4 and 5.
This is compatible with our best-fit SEDs: for epoch 4, the extrapolation
of the GROND optical/NIR SED nearly exactly reproduces the APEX/LABOCA
measurement, while for epoch 5 the optical/NIR flux (determined 
from an interpolation between two GROND measurements) has faded more rapidly
than the sub-mm flux, resulting in a move of $\nu_m$ to lower
frequencies. The speed of this frequency 
displacement between epoch 4 and 5 is
measured as $t^{-1.15\pm0.55}$, 
consistent with the fireball prediction of $t^{-3/2}$.
The observed optical/NIR flux is about a factor 2 above the extrapolation
of the decay of the canonical afterglow, thus we assign this emission
to the jump component.
In contrast, as shown in the previous sub-section, the X-ray
emission is due to the canonical afterglow component. Curiously, 
despite the steeper X-ray spectrum, a
formal SED fit including the X-rays is possible due to the large
gap between the optical and X-ray bands: since the X-ray spectrum
has a steeper slope than the optical/NIR/sub-mm at this time,
the large allowed range for $\nu_c$ can accommodate this bright
X-ray component.

Thus, with the two
assumptions that
(i) the contemporaneously measured X-ray emission is a separate
emission component, and therefore is left out from fitting; and
(ii) the long-wavelength part is dominated by the jump component,
we make a combined spectral fit of epochs 3--6, where only
epoch 3 contains X-ray data. We fix $\beta_1 = 1/3$, $\Delta\beta = 0.5$
between the GROND and the \textit{Swift}/XRT band,
and also fix the host extinction at the value of $A_{\rm V} = 3.65$ mag as
derived from the fit of epoch 3. With the lower S/N ratio of the later
GROND SEDs, the slope in the GROND range is largely dominated by epoch 3,
with a best-fit value for the combined fit of $\beta_2 = 0.90\pm0.04$.
The ATCA measurements then define the break energy $\nu_m$
as summarized in Tab. \ref{breaks}.
We note that a fireball-compliant evolution of $\nu_m$ from these
values extrapolated backwards in time does not conflict the limit on $\nu_m$
set by the NIR data at 5.5 ks (see dashed line labeled ``$t^{-3/2}$'' in 
Fig. \ref{freqmove}).

\subsubsection{Epochs 6 and 7}

For these two epochs, we have the radio fluxes at two frequencies 
from the ATCA measurements. As described earlier, they are compatible
with the $\nu^{1/3}$ slope as expected for the segment between $\nu_{sa}$
and $\nu_m$. At sub-mm, the APEX/LABOCA upper limit is well above
this spectral component, and does not constrain the SED.

The more or less unchanged radio flux in epochs 6 and 7 (formal fit
results in $t^{-0.5}$, though the large error bars of epoch 7 
also allow a slightly rising flux) implies that the injection frequency
is in the few GHz range (near our radio data). 
This conclusion is supported by two other
observational constraints, namely that the radio spectral slope is 
somewhat flatter than $\nu^{1/3}$, and that the radio flux must decline
within the following 20 days in order to be compatible with the
ATCA upper limits (see Tab. \ref{ATCA}).

\begin{figure}[th]
\hspace{-0.5cm}\includegraphics[width=9.8cm]{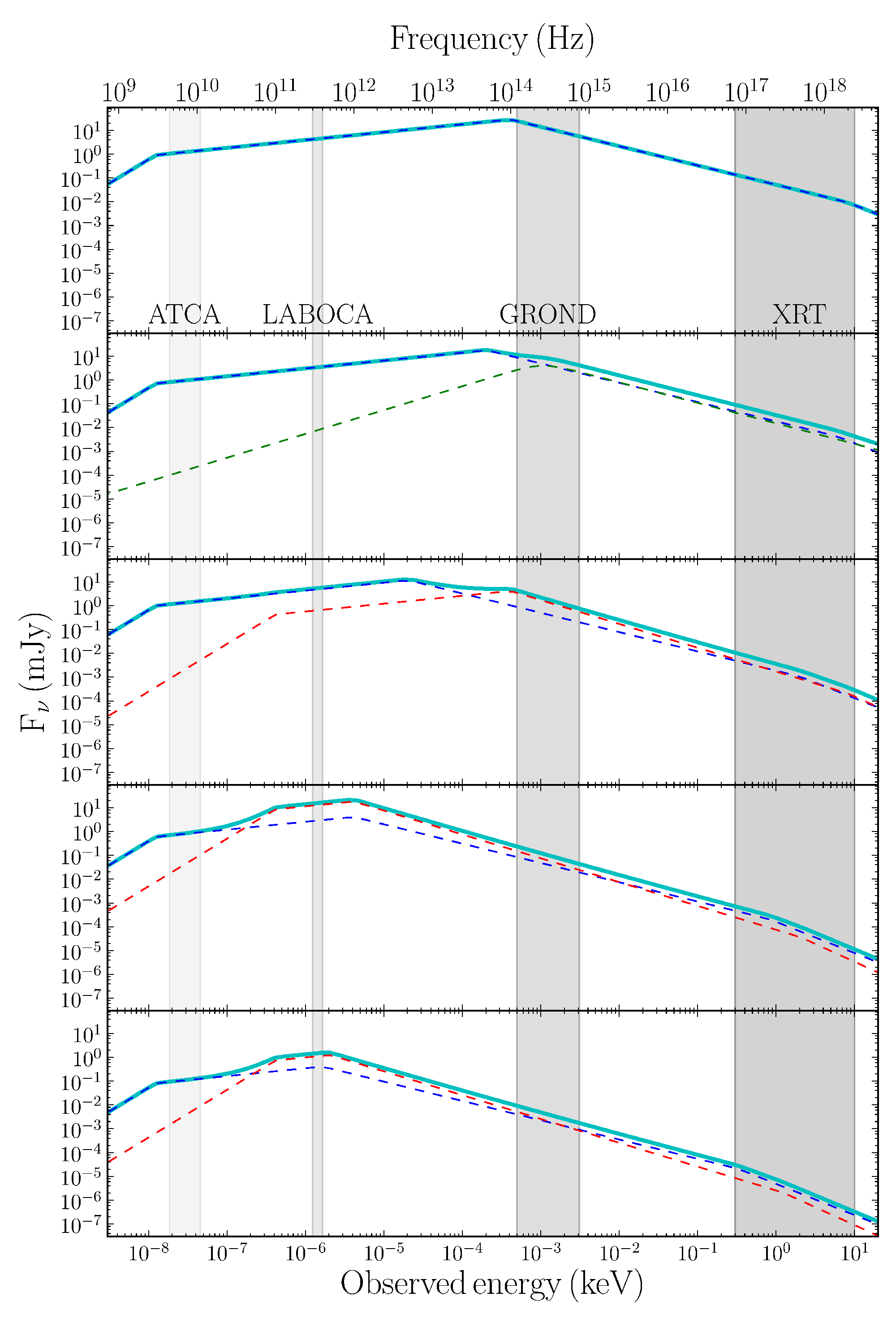}
\caption{Visualization of the spectral energy distributions at the
five epochs as discussed in the text, with panel 1 to 3 showing epochs
1-3, panel 4 showing epoch 4/5, and panel 5 showing epoch 6/7.
The frequency/energy ranges covered by our observations 
are marked as shaded bands. Dashed lines mark the different emission
components: afterglow (blue), flares (green), jump component (red).
The thick line is the sum of these components.
\label{scetch}
}
\end{figure}

A combined fit of the radio and optical/NIR data results in a 
best-fit injection frequency of order 2$\times 10^{12}$ Hz, a factor 1000
larger than our above estimate, and also a factor 10 larger than one
would expect from the (fireball-compliant) evolution of the jump component.
This suggests that the
radio emission belongs to the canonical afterglow component, 
while the optical/NIR belongs to the jump component (as argued above).
This picture is consistent with a (again fireball-compliant) 
prediction of the early evolution of $\nu_m$, i.e. that 
$\nu_m$ is at frequencies shortward of the GROND NIR measurements
at very early times (see the blue dashed line in Fig. \ref{freqmove}).

\subsection{Characterization of the three emission components}

First, Tab. \ref{compo} summarizes the discussion from the above sub-sections
with respect to the three emission components, rather than according
to the epoch of observation, and Fig. \ref{scetch} provides a 
visualization of the evolution of  these three components with time.

\begin{table}[ht]
  \caption[]{Epochs at which the three emission components are seen 
    at different frequency bands.
  \label{compo}}
   \vspace{-0.2cm}
   \begin{tabular}{cccc}
   \hline
   \noalign{\smallskip}
             & canonical afterglow          & flares    & jump component \\
   \noalign{\smallskip}
   \hline
   \noalign{\smallskip}
 X-rays      & 1, 3: partially, 4-7: fully  & 2         &   3: partially \\
 optical/NIR & 1: partially                 & 2         & 3-7: fully \\
 sub-mm      &  --                          & --        & 4+5: fully \\
 radio       & 6+7: fully                   & --        & -- \\
   \noalign{\smallskip}
   \hline
    \noalign{\smallskip}
  \end{tabular}
\end{table}

With these constraints on the varying combination of the three emission 
components at a given epoch, 
the combined fitting results in a total reduced
$\chi^2_{\rm red} = 1.1$ (162 for 145 degrees of freedom), 
thus being an acceptable fit. The best fit
power law slope in the GROND range for the canonical afterglow 
(fully described in section 4.2.1) is $\beta_2$ = 0.82$\pm$0.02
with a strong host extinction of A$_{\rm V}$ = 3.65$\pm$0.06 mag,
as already indicated by the very red colors of the afterglow.

With the generic picture that the typical afterglow spectrum evolves 
from fast to slow cooling, 
we will now use the constraints for each of the components, 
and try to infer a consistent picture of the evolution
of the GRB 100621A afterglow. The discussion is based on the formalism
described in \citet{grs02}, and we use the same nomenclature of
$E_{52} = E / 10^{52}$ erg, and 
~$\epse = \epsilon_e (p-2)/(p-1)$.

\subsubsection{The canonical afterglow}

The SED of epoch 1 provides four constraints on the fireball parameters
of the canonical afterglow:
(i) a lower limit on the frequency of $\nu_c$ at that time ($>$8 keV), 
(ii) an upper limit on the flux density at  $\nu_c$ ($<$0.035 mJy),
(iii) an upper limit on $\nu_m$ based on the non-detection of $\nu_m$
(or a break in general) in the GROND range, 
i.e. $\nu_m < 1.25\times 10^{14}$ Hz  ($<$2.4 $\mu$m), and 
(iv) a lower limit on the flux at this limit frequency ($>$9 mJy). 
Using the two equations each in lines
3 and 5 of Tab. 2 of \cite{grs02}, these measurements translate
into the following four conditions:

\begin{equation}
\begin{array}{l}
{\rm ~~(i)} \hspace{1.cm} \epsilon_B^{-3/2} \cdot n^{-1} \cdot E_{52}^{-1/2} > 2.82\times10^4 \\

{\rm ~(ii)}\hspace{1.cm} \epse^{1.62} \cdot \epsilon_B^{2.12} \cdot n^{1.31} \cdot
     E_{52}^{1.81} < 7.17\times10^{-9} \\

{\rm (iii)} \hspace{1.cm} \epse^2 \cdot \epsilon_B^{1/2} \cdot E_{52}^{1/2} < 6.46\times10^{-6} \\

{\rm (iv)} \hspace{1.cm} \epsilon_B^{1/2} \cdot n^{1/2} \cdot E_{52} > 0.195 \\

\end{array}
\end{equation}

\begin{figure}[ht]
\hspace{-0.2cm}\includegraphics[angle=270,width=9.6cm]{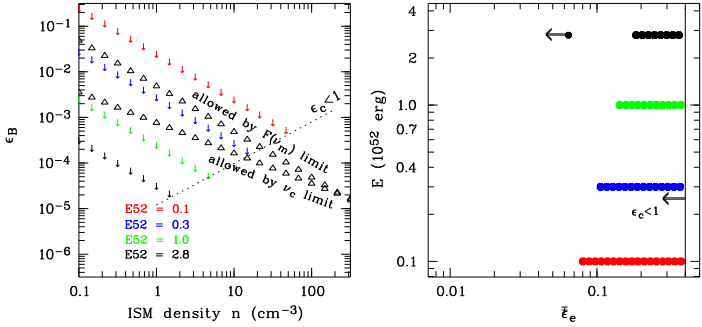}
\vspace{-1.8cm}
\caption{Constraints on the microphysical parameters of the canonical 
afterglow component from the SED of epoch 1 
at 520 s after the GRB (black triangles), and of epoch 6 
(colored arrows/lines).
The limits on $F(\nu_m)$ and $\nu_c$ from epoch 1 allow
the parameter space \emph{above} the lines of open triangles (left panel),
and an upper limit of \,$\epse < 0.064$ (top arrow on right panel).
The limits from epoch 6/7 allow the parameter space \emph{below}
the lines of arrows (left panel), depending on the total energy.
The requirement that $\epsilon_e < 1$ translates into an upper limit
in the density and lower limit on $\epsilon_B$, respectively (dotted line).
The thick colored lines on the right panel show the corresponding
allowed range for \,$\epse$ 
(where  \,$\epse = \epsilon_e \times (p-2)/(p-1) = 0.39 \epsilon_e$ for
the derived p=2.64).
\label{agpar}
}
\end{figure}

These equations define an upper limit on \,$\epse < 0.064$ (which translates
into $\epsilon_e < 0.16$ for our $p$=2.64), and combined lower limits for
$\epsilon_B$ and the external density as shown by the lines of arrows
in Fig. \ref{agpar}.

In principle, there are two more constraints, 
namely the limit that the time of $\nu_m$ crossing $\nu_c$  
($5\rightarrow 1$ in \citep{grs02}) has occured within $<$520 s, 
and the  transition $1 \rightarrow 2$ ($\nu_m$ crossing $\nu_{sa}$) 
is constrained to $>$416 ks. 
However, these limits do not impose any additional constraints 
as  shown in Fig. \ref{agpar}.

Epoch 3 does not provide any further constraint on the canonical afterglow,
as the X-ray flux and spectral shape cannot be independently differentiated
from that of the jump component, as mentioned above.

At epochs 4/5, the X-rays provide the only measurements of the
canonical afterglow. Given the somewhat contrived conclusion that
the late X-ray light curve after the
break at 80 ks is due to a combination of cessation of energy injection and
cooling break passage, {\emph and} that it is still in the spherical phase, 
we refrain from adding these constraints here.

Epochs 6/7, after re-fitting without the GROND optical/NIR data, 
provide no unambiguous measurement of $\nu_m$ and $F(\nu_m)$ 
(or $\nu_c$), as the normalization of the power law segment
which connects the $\nu^{-1/3}$ segment with the X-ray segment, 
is not constrained. Stepping through $\nu_m$ in the range 
1$\times$10$^{-7}$ keV to 6$\times$10$^{-5}$ keV reveals equally
good fits as long as $\nu_m > 5 \times 10^{-6}$ keV (250 $\mu$m).
With this limit, we obtain:
\begin{equation}
\begin{array}{l}

{\rm ~~(i)} \hspace{1.cm} \epse^2 \cdot \epsilon_B^{1/2} \cdot E_{52}^{1/2} > 1.1\times10^{-3} \\

{\rm ~(ii)} \hspace{1.cm} \epsilon_B^{1/2} \cdot n^{1/2} \cdot E_{52} < 0.017 \\

\end{array}
\end{equation}

\noindent All combined constraints for the afterglow component are shown in 
Fig. \ref{agpar}.

\subsubsection{The early flares}

The  flares are only observed at early times, and for a description we
have picked epoch 2 to cover some of those.
While we called these events flares, it seems obvious that these
are somewhat dissimilar to the canonical X-ray flares observed
by \textit{Swift}/XRT in a large fraction of GRBs: in the case of GRB 100621A,
the flares are prominent in the optical, rather than in X-rays.
If these have the same origin as the canonical X-ray flares
\citep{mbb11}, the only difference might be a lower peak energy.
The broad-band spectrum between GROND and \textit{Swift}/XRT is certainly
not a single power law (see section 4.2.1). As the low-energy
part of a broken power law fit ($\beta = 0.86$) would be very steep
for a Band function approach, the peak energy $E_{peak}$ rather is
below the GROND wavelengths, with possibly some exponential cut-off
at X-rays. Since the decomposition of normal afterglow component
and flares is not unique, no statement can be made on a possible
variation of the peak energy with time.

\subsubsection{The jump component}

\begin{figure}[ht]
\hspace{-0.2cm}\includegraphics[angle=270,width=9.6cm]{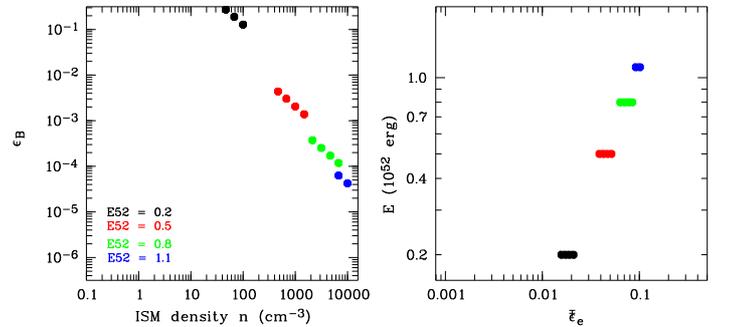}
\vspace{-2.0cm}
\caption{Constraints on the microphysical parameters of the jump component,
as derived from epoch 4/5 and 6/7 (see text) for the case of a constant 
ISM density profile.
\label{jumppar}}
\end{figure}

As mentioned earlier, the \cite{vlasis11} interpretation of the
optical/NIR emission at 5--8 ks is via the collision of two 
ultrarelativistic shells.

The SED of the 5--8 ks event 
exhibits a straight power law of slope 1.1$\pm$0.1
covering the GROND optical/NIR and the \textit{Swift}/XRT region.
If interpreted using the \cite{grs02} formalism for afterglows
(the applicability of which is not obvious as the medium into which
the colliding shell is evolving might be increasing in density, rather
than being constant or decreasing)
the location of $\nu_{c}$ and $\nu_{m}$ remain ambiguous.
If $\nu_{c}$ were longwards of 2.4 $\mu$m (GROND $K_{s}$-band),
then the electron spectral index would be a reasonable $p$=2.2.
However, in addition to the observed light curve decay at $>$10 ks being much
steeper than the expected $t^{-1.15}$, there would be further 
inconsistencies: (1) if the circumburst environment had a wind
density profile, $\nu_{c}$ would evolve to higher frequencies,
i.e. into the GROND band, which is not observed; (2) if, alternatively,
the circumburst density profile were ISM-like, $\nu_{c}$ would
move towards the LABOCA band. However, at epoch 4 the optical/NIR
SED extrapolates nearly perfectly to the measured LABOCA flux,
thereby not allowing any break. Thus, $\nu_{c}$ would have to be
below 345 GHz at epoch 3. This would imply a later radio flux
at least a factor 10$^3$ larger than observed, and therefore can be
excluded. We thus conclude that $\nu_{c}$ at epoch 3 must be $>$8 keV,
implying a steep $p$=3.2. 
Using these constraints, we arrive at the following four conditions:

\begin{equation}
\begin{array}{l}
{\rm ~~(i)} \hspace{1.cm} \epsilon_B^{-3/2} \cdot n^{-1} \cdot E_{52}^{-1/2} > 1.58\times10^5 \\

{\rm ~(ii)}\hspace{1.cm} \epse^{2.2} \cdot \epsilon_B^{-1/2} \cdot n^{1.6} \cdot
     E_{52}^{2.1} < 2.47\times10^{-10} \\

{\rm (iii)} \hspace{1.cm} \epse^2 \cdot \epsilon_B^{1/2} \cdot E_{52}^{1/2} < 2.36\times10^{-4} \\

{\rm (iv)} \hspace{1.cm} \epsilon_B^{1/2} \cdot n^{1/2} \cdot E_{52} > 0.358 \\

\end{array}
\end{equation}

\noindent None of these conditions is violated by the constraints derived below
for the emission of the 5--8 ks event.

The SED of this 5--8 ks event is constrained by our
measurements of epochs 4/5 and 6/7. During epoch 4, we measure
$\nu_m$ and $F(\nu_m)$, which provides the following two equations:
\begin{equation}
\begin{array}{l}

{\rm ~~(i)} \hspace{1.cm} \epse^2 \cdot \epsilon_B^{1/2} \cdot E_{52}^{1/2} = 7.0\times10^{-5} \\

{\rm ~(ii)} \hspace{1.cm} \epsilon_B^{1/2} \cdot n^{1/2} \cdot E_{52} = 0.716 \\

\end{array}
\end{equation}

Epochs 6/7 provide an interesting constraint on the sub-mm/radio regime,
despite the non-detections longward of the GROND-$K$ band. The APEX/LABOCA
non-detection does not constrain the continuation of the optical/NIR slope
into the mm-band, but a fireball-compliant extrapolation would suggest
$\nu_m \approx 1\times10^{11}$ Hz at epoch 6. Since we have argued earlier
that the radio emission seen at this epoch at 5.5 and 9 GHz must belong 
to the canonical afterglow, we have to assume that the radio-component
of the jump component must be self-absorbed to a level to not exceed
the measured fluxes at 5.5 and 9 GHz. This results in 
$\nu_{sa}$ \gax $0.8\times10^{11}$ Hz, i.e. $\nu_m = \nu_{sa}$ at epoch 6
(and 7) to within the errors. This is exactly what \cite{vlasis11}
find during the modelling of the radio light curve: the amplitude
is strongly depressed due to self-absorption.

In a constant external density profile, $\nu_{sa}$ is constant, and our
above assumption does not violate any observational constraint at
earlier or later times. For a wind environment, $\nu_{sa}$ decreases
according to $t^{-3/5}$ -- this is slow enough that it does not conflict
with the LABOCA detections at epochs 4/5.

Thus, for the ISM case, we derive:

\begin{equation}
\begin{array}{l}

{\rm (iii)} \hspace{1.cm} \epse~^{-1} \cdot \epsilon_B^{1/5} \cdot n^{3/5} \cdot E_{52}^{1/5} = 538 \\

{\rm (iv)} \hspace{1.cm} \epse~^{-1} \cdot \epsilon_B^{2/5} \cdot n^{7/10} \cdot E_{52}^{9/10} > 123 \\

\end{array}
\end{equation}

\noindent The combination of the last 4 equations translates into 
the two thin stripes of parameter space shown in Fig. \ref{jumppar}.
The resulting limits on the
external density are rather high: since $\epsilon_B$ cannot be larger
than 1, the external density must be \gax 20 cm$^{-3}$. Moreover,
the total energy is constrained to $E_{52} > 0.2$, and \,$\epse > 0.01$.
We stress again that these constraints are only valid if the
\cite{grs02} formalism is applicable.

\section{Discussion}

\subsection{Fitting assumptions and results}
 
The behaviour of the afterglow of GRB 100621A at different epochs and
frequencies has been found to be too complex relative to our set of
observational data to be able to constrain models. We therefore have
adopted the fireball scenario and attempted to construct a consistent
picture of the observed features. Before further discussion, we summarize
our assumptions here:
(i) First, we assume that the total emission is due to the superposition 
of 3 emission components;
(ii) we have fixed $\Delta {\beta}$=0.5 between X-ray and GROND power law slopes
(whenever applicable);
(iii) we have fixed $\beta_{radio}$ = $-$1/3 as derived from the two radio 
frequencies at epochs 6/7; and
(iv) had to assume that the the jump component has to be self-absorbed 
in the radio.
With these assumptions, we find a reasonably consistent picture which 
describes all of our observational facts (temporal and spectral slopes) 
except the slow X-ray decay at times $>$80 ks.

For none of the three emission components in the afterglow of GRB 100621A
do we have enough observations at the right time to determine all
fireball model parameters in a unique way. The constraints on these
parameters as derived from our observations are, in general, broadly
consistent with expectations.
The only inconsistent result is that for $\epsilon_B$ of the afterglow 
component: the lower limits from epoch 1 are about 2 orders of magnitude
higher than the upper limits as derived from epochs 6/7, assuming otherwise
equal parameters (in particular total energy and density). There could
be several reasons for this, one of which could be an evolving
$\epsilon_B$ with time, though we do not consider this.
A more obvious reason could be that the energy
ejection (which was deduced to make spectral and temporal slopes in the
early phases consistent with the fireball scenario) introduces a 
time-dependent variation between low- and high-frequency segments
(at radio wavelength, the impact of the energy injection will come
later than at X-rays). This invalidates our assumption for epochs 6/7
in deriving constraints on $\nu_m$, in that the radio and X-ray sections
of the SED reflect the same internal energy budget. We therefore neglect
the $\nu_m$ constraints from epochs 6/7 in the following.
If we allow $\nu_m$ to be just above 9 GHz during epochs 6/7, then
no conflicting constraints are imposed anymore.

\begin{figure}[ht]
\includegraphics[angle=270,width=8.8cm]{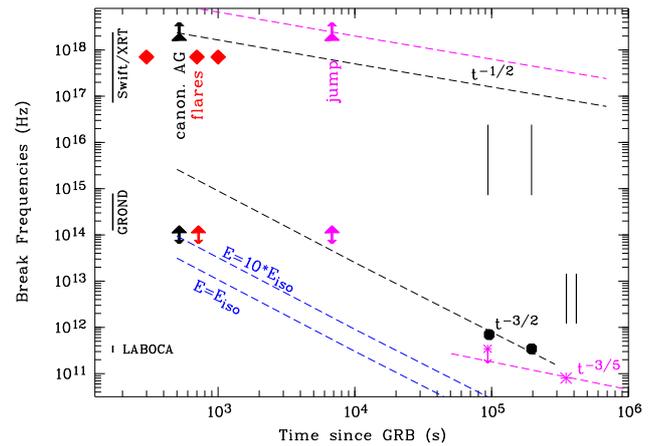}
\vspace{-0.2cm}
\caption{Location of the two breaks $\nu_c$ (top end) and $\nu_m$ 
(bottom part) at different epochs in the late-time evolution of the 
afterglow of GRB 100621A, for each of the three emission components 
(i) canonical afterglow (black), 
(ii) flares (red), and (iii) jump component (pink).
Vertical bars indicate allowed ranges for $\nu_c$ or  $\nu_m$.
The wavelength coverage of our instruments is shown as vertical bars
at the very left side. 
Dashed lines show the expected evolution according to the standard
fireball scenario after obeying limits as derived from our observations
at various epochs.
\label{freqmove}
}
\end{figure}

Despite the complex behaviour, we are able to unequivocally
deduce a constant ISM-like circumburst density profile. The slow
intensity decline of the external forward shock suggests
continuous energy injection at a rate proportional to $t^{0.35}$
during the first hour after the GRB.
With the onset of the jump component, another sudden increase in
energy happens which lifts the energy budget by a factor 2--5.

One could imagine that the canonical afterglow and the early flares 
experience the same external ISM density, i.e. that they originate co-spatially.
In this case, the combined constraints imply that the external density
$n$ \gax 50 cm$^{-3}$, otherwise the $F(\nu_m$) limit for the afterglow
component would be violated. This in turn would imply that the energy
driving the flares would be of order $E_{iso}$ ($1 < E_{52} < 5$), 
which is surprisingly large though not exceptional.
Correspondingly, we deduce $0.014 < \epse < 0.064$ and $\epsilon_B > 10^{-4}$.

For the jump component, as mentioned above, we derived $n$ \gax 20 cm$^{-3}$.
This is interesting as one could have imagined that this component
originates in the wake of the afterglow, i.e. in a region 
cleared by the forward shock. However, we caution (again) that
the interpretation with the \cite{grs02} framework might not be
appropriate at all. Further theoretical investigation of such
shell collisions are certainly warranted.

\subsection{Location of the dust}

From multiple SED fits during the early rise and early plateau 
(around 200--400 s after the GRB trigger) we constrain any variation 
of the extinction to $\Delta A_{\rm V} <$ 10\%. The intense radiation
of gamma-ray bursts has been repeatedly suggested to destroy the dust
in its near environment through sublimation \citep{wad00, fkr01, pel02},
out to distances of a dozen parsec. The large dust column we observe
in the afterglow of GRB 100621A must therefore be at larger distances,
most likely not related to the star formation site of the progenitor
of GRB 100621A.

\begin{figure}[ht]
\includegraphics[angle=270,width=9.0cm]{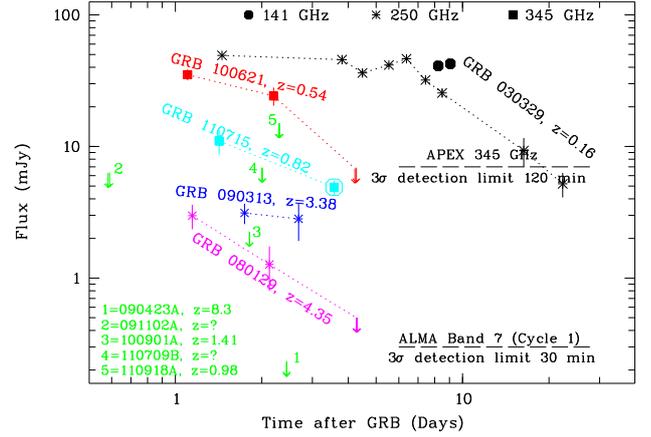}
\vspace{-0.2cm}
\caption{Comparison of our GRB 100621A sub-mm light curve to previous
sub-mm observations of GRBs with more than one observation, and selected
upper limits for a few famous GRBs. 
Different symbols mark different observer frequencies, and colors
denote different GRBs (except for the upper limits).
Data are from:
 GRB 030329: \citep{ktt05, sheth03};
 GRB 090313: Greiner et al 2013, in prep;
 GRB 080129: \citep{gkm09};
 GRB 090423: \citep{bcf09};
 GRBs 091102, 110709B, 110715A, 100901A, 110918A: \citep{ulm12}.
}
\label{submmlc}
\end{figure}

\subsection{Comparison with previous sub-mm detections}

Previous sub-mm measurements of GRB afterglows were initially
non-detections \citep{bremer98, shepard98}, and detections or
even light curves are sparse \citep{cbr08, sheth03, gkm09, pah12, zlb12}. 
Predictions of emission at flux levels of several tens of mJy 
\cite[e.g.][]{inoue05} have not been materialized.
So far, only a handful of GRBs have been detected in the mm/sub-mm, 
mostly using MAMBO at the IRAM 30m 
\citep{cbr08, sheth03, gkm09}, and CARMA \citep{cbs07, bcf09, pah12}.
GRB 100621A is one of a handful of GRBs for which a sub-mm 'light curve'
(more than 1 detection) is available (Fig. \ref{submmlc}).
However, the complicated early optical/NIR light curve
of GRB 100621A makes even this relatively well-observed GRB too sparsely
sampled in the sub-mm range, which leaves ambiguities in the interpretation
of both, the light curve and the movement of the low-frequency break.


Recent more aggressive attempts with APEX/LABOCA have confirmed to return
mostly non-detections \citep{ulm12}, indicating that the injection frequency
moves rather rapidly to frequencies below the LABOCA range, thus 
requiring sub-mm observations within the first day in order to 
achieve detections. APEX/LABOCA is able to do this for the best suited
afterglows (steep optical/NIR SED), but for the majority ALMA will
be the instrument of choice, once rapid turn-around target-of-opportunity
observations will be offered.

\subsection{The GRB host}

The host galaxy of GRB 100621A was extensively covered in \cite{kgs11},
including in addition to the GROND and 
\textit{Swift}/UVOT data.
In short, the r'$\approx$21.5 mag galaxy is well detected from 
the UV (all \textit{Swift}/UVOT filters) up to the $K_{s}$-band showing a very blue
spectral energy distribution with ($R-K$)$_{\rm AB} \approx 0.3$ mag.
The stellar population synthesis fitting of the host SED returns 
an age of the dominating stellar population of only 0.05 Gyr, and an intrinsic
extinction of A$_{\rm V}^{host}$ = 0.6$^{+0.1}_{-0.2}$ mag, 
in stark contrast to the large  afterglow (AG) extinction of
A$_{\rm V}^{AG}$ = 3.61$\pm$0.06 mag. The absolute magnitude of the host
is M$_{\rm B} = -20.68\pm0.08$ mag, and the star formation rate 
was determined as 13$^{+6}_{-5}$ \msun/yr.

The APEX and ATCA non-detections of any flux at the position of GRB 100621A
at $>$5 days after the GRB also provide first crude limits on the
sub-mm and radio emission of the host galaxy, of $<$6.8 mJy at 345 GHz,
$<$170 $\mu$Jy at 5.5 GHz, and $<$200 $\mu$Jy at 9 GHz 
(all 2$\sigma$ confidence). 
Assuming that the dominant fraction of the radio emission would be of
non-thermal origin, and using the formalism of \cite{yuc02}, this 
implies an upper limit on the star formation rate of \lax 100 \msun/yr.

Due to the bright, compact host, no observational attempt has been made 
with GROND to
search for the supernova component which would have peaked about 6 magnitudes
fainter (if extinguished the same way as the afterglow)
than the host brightness for a 1998bw-like SN-luminosity.

\section{Conclusions}

GRB 100621A has shown the brightest X-ray emission after any gamma-ray 
bursts so far. Despite this, the afterglow at \gax 200 s was not
extraordinarily bright, and the strong host extinction made it only
marginally detectable in \textit{Swift}/UVOT observations. Yet, we obtained
a decent data set with GROND as well as supporting APEX/LABOCA
and ATCA measurements. 

The biggest surprise in the properties of the afterglow of GRB 100621A
is undoubtly the sudden intensity jump after about 1 hr. Here, we have been
able to characterize its properties in hitherto unprecedented detail. The 
pecularity of this event is the complexity of the combined afterglow
emission which we encounter. In order to disentangle this complexity,
and to possibly even test afterglow models, a much denser sampling
of the afterglow emission in time is required, both at sub-mm as well
as radio frequencies. At least for sub-mm observations from the southern
hemisphere, ALMA would be an ideal instrument if fast reaction times
to external alerts like gamma-ray bursts can be implemented.

\begin{acknowledgement}
JG expresses special thanks to A. Vlasis for discussing some details
of the shell collision scenario.
We are grateful to ESO for approving the DDT proposal for APEX observations.
Particular thanks to A. Kaufer for the support in the scheduling discussions
for technical and Chilean time.
AM was a fully sponsored PhD candidate at ICRAR - Curtin University until 
2011 and acknowledges the support of SHAO as a postdoctoral research fellow.
We are similarly grateful to P. Edwards for approving and scheduling
the ATCA ToO and regular observations.
TK acknowledges support by
the DFG cluster of excellence 'Origin and Structure of the Universe'
during the early part of this project when being employed at MPE,
and AU is grateful for travel funding support through MPE.
FOE acknowledges funding of his Ph.D. through the 
\emph{Deutscher Akademischer Austausch-Dienst} (DAAD),
SK and ARossi acknowledge support by DFG grant Kl 766/13-2,
and SK, ARossi, ANG and DAK acknowledge support by DFG grant Kl 766/16-1.
ARossi additionally acknowledges support from the BLANCEFLOR 
Boncompagni-Ludovisi, n\'ee Bildt foundation, and through 
the Jenaer Graduierten\-akademie.
MN acknowledges support by DFG grant SA 2001/2-1.
Part of the funding for GROND (both hardware as well as personnel)
was generously granted from the Leibniz-Prize to Prof. G. Hasinger
(DFG grant HA 1850/28-1).
The Dark Cosmology Center (TK) is funded by the Danish National Research 
Foundation.
This work made use of data supplied by the UK \textit{Swift} Science Data 
Centre at the University of Leicester.
\end{acknowledgement}

\bigskip

\noindent {\small {\it Facilities:} Max Planck:2.2m (GROND), 
                  Swift}


\begin{thebibliography}{}

\bibitem[Beuermann et al.(1999)]{bhr99} Beuermann K., Hessman F.V., 
  Reinsch K. et al.\ 1999, \aap, 352, L26

\bibitem[Blustin et al.(2006)]{bbb06} Blustin A.J., Band D., Barthelmy S.
  et al. 2006, ApJ 637, 901

\bibitem[Bock et al.(2009)]{bcf09} Bock D.C.-J., Chandra P., Frail D.A., 
  Kulkarni S.R., 2009, GCN \#9274

\bibitem[Bremer et al.(1998)]{bremer98} Bremer M., Krichbaum T.P., Galama T.J.
et al. 1998, A\&A, 332, L13

\bibitem[Chandra et al.(2007)]{cbs07} Chandra P., Bock D., Soderberg A.
et al. 2007, GCN 6073

\bibitem[Chandra et al.(2008)]{cbr08} 
Chandra P., Cenko S.B., Frail D.A.  
et al.  2008, ApJ 683, 924

\bibitem[de Ugarte Postigo et al.(2012)]{ulm12} de Ugarte Postigo A., 
Lundgren A., Martin S., et al. 2012,  A\&A 538, 44D

\bibitem[Evans et al.(2007)]{ebp07} 
Evans, P.A., Beardmore, A.P., Page, K.L.,
   et al. 2007, A\&A, 469, 379
   
\bibitem[Evans et al.(2009)]{ebp09} 
Evans, P.A., Beardmore, A.P., Page, K.L.,
   et al. 2009, MNRAS, 397, 1177

\bibitem[Evans et al.(2010)]{ego10}  Evans, P.A., Goad M.R., Osborne J.P., 
  Beardmore, A.P., 2010, GCN \#10873

\bibitem[Filgas et al.(2012)]{fgs12} Filgas R., Greiner J., Schady P. 
 et al. 2012, A\&A 535, A57

\bibitem[Fruchter et al.(2001)]{fkr01} Fruchter A., Krolik J.H., Rhoads J.E., 
2001, ApJ 563, 597

\bibitem[Gehrels et al.(2004)]{gcg04} Gehrels N., Chincarini G., Giommi P., 
et al. 2004,    ApJ 621, 558

\bibitem[Ghirlanda et al.(2012)]{gng12} Ghirlanda G., Nava L., Ghisellini G.
  et al. 2012, MN 420, 483

\bibitem[Golenetskii et al.(2010)]{gaf10} Golenetskii S., Aptekar R., 
  Frederiks D. et al. 2010, GCN \#10882

\bibitem[Granot et al.(1999)]{gps99} Granot J., Piran T., Sari R., 1999,
ApJ 527, 236

\bibitem[Granot \& Sari(2002)]{grs02} Granot J., Sari R., 2002, ApJ 568, 820

\bibitem[Greiner et al.(2008a)]{gbc08}
Greiner J., Bornemann W., Clemens C., et al. 2008a, PASP 120, 405

\bibitem[Greiner et al.(2009a)]{gkf09} Greiner J., Kr\"uhler T., Fynbo J.P.U, 
et al. 2009a, ApJ 693, 1610

\bibitem[Greiner et al.(2009)]{gkm09} Greiner J., Kr\"uhler T., McBreen S., 
et al. 2009, ApJ 693, 1912

\bibitem[Greiner et al.(2011)]{gkk11} 
Greiner J.,  Kr\"uhler T., Klose S. et al. 2011, A\&A 526, A30

\bibitem[Greiner(2011)]{grei11} Greiner J., 2011, talk at ``The prompt 
activity of Gamma-Ray Bursts'', Rayleigh, March 2011,
http://grb.physics.ncsu.edu/GRB\_2011/WEB/TALKS/greiner.pdf

\bibitem[Inoue et al.(2005)]{inoue05} Inoue S., Omukai K., Ciardi B., 2005, 
MN 380, 1715

\bibitem[Kohno et al.(2005)]{ktt05} Kohno K., Tosaki T. Okuda T. et al. 2005, 
PASJ 57, 147

\bibitem[Kov\'acs(2008)]{kov08} 
Kov\'acs A., 2008, Proc. SPIE vol. 7020, id. 70201S-15

\bibitem[Kr\"uhler et al.(2008)]{kkg08} Kr\"uhler T., K\"{u}pc\"{u} Yolda\c{s}
 A., Greiner J.,   et al. 2008, ApJ 685, 376

\bibitem[Kr\"uhler et al.(2011a)]{ksg11} Kr\"uhler T., Schady P., Greiner J., 
et al. 2011a, A\&A 526, A153

\bibitem[Kr\"uhler et al.(2011b)]{kgs11} Kr\"uhler T., Greiner J., Schady P. 
et al. 2011b, A\&A 534, A108

\bibitem[K\"{u}pc\"{u} Yolda\c{s} et al.(2008b)]{kkg08b} K\"{u}pc\"{u} Yolda\c{s} A., Kr\"uhler T., Greiner J., 
 et al. 2008b, AIP Conf. Proc., 1000, 227

\bibitem[Margutti et al.(2011)]{mbb11} Margutti R., Bernardini G., 
Barniol Duran R.,
2011, MN 410, 1064

\bibitem[Meszaros \& Rees(1997)]{mer97} Meszaros P., Rees M.J., 
1997, ApJ 476, 232

\bibitem[Meszaros et al.(1998)]{mes98} M\'{e}sz\'{a}ros P., Rees M.J., 
Wijers R.A.M.J., 1998, ApJ 499, 301 

\bibitem[Milvang-Jensen et al.(2010)]{mgt10} Milvang-Jensen B., Goldoni P., 
  Tanvir N.R. et al. 2010, GCN \#10876

\bibitem[(Molinari et al. 2007)]{mol07}
Molinari E., et al. 2007, A\&a 469, L13

\bibitem[Naito et al.(2010)]{nss10} Naito H., Sako T., Suzuki D., 
et al. 2010, GCN \#10881

\bibitem[Nardini et al.(2011)]{ngk11} Nardini M., Greiner J., Kr\"uhler T. 
 et al. 2011,   A\&A 531, A39

\bibitem[Nousek et al.(2006)]{nkg06} Nousek J.A., Kouveliotou C., Grupe D. 
et al. 2006, ApJ 642, 389

\bibitem[Panaitescu et al.(2006)]{pmg06} Panaitescu A., Meszaros P., Gehrels N.,
  et al. 2006, MN 366, 1366

\bibitem[Panaitescu \& Vestrand(2008)]{pav08} Panaitescu A., Vestrand W.T.,
2008, MN 387, 497

\bibitem[Pandey et al.(2010)]{prg10} Pandey S.B., Rujopakarn W., Guver T., 
et al. 2010, GCN \#10871

\bibitem[Pei(1992)]{pei92} Pei Y.C. 1992, ApJ 395, 130

\bibitem[Perley et al.(2012)]{pah12} Perley D.A., Alatalo K., Horesh A., 
2012, GCN 13175

\bibitem[Perna \& Lazzati(2002)]{pel02} Perna R., Lazzati D., 2002, ApJ 580, 261

\bibitem[Racusin et al.(2008)]{rac09} Racusin J.L., Karpov S.V., Sokolowski M.
et al. 2008, Nat. 455, 183

\bibitem[Sari et al.(1998)]{sari98} Sari R., Piran T., Narayan R., 
1998, ApJ 497, L17

\bibitem[Sari(1999)]{sari99} Sari R., 1999, ApJ 524, L43

\bibitem[Schlegel et al.(1998)]{sfd98} Schlegel D., Finkbeiner D., Davis
 M. 1998, ApJ 500, 525

\bibitem[Schuller(2012)]{schu12} 
Schuller F., 2012, Proc. of SPIE vol. 8452, id. 84521T-10

\bibitem[Shephard et al.(1998)]{shepard98} Shepherd D.S., Frail D.A., 
Kulkarni S.R., \& Metzger M.R., 1998, ApJ 497, 859

\bibitem[Sheth et al.(2003)]{sheth03} Sheth et al. 2003, ApJ 595, 33

\bibitem[Siringo et al.(2009)]{siringo09} Siringo G., Kreysa E., Kovacs A., 
et al. 2009, A\&A 497, 945

\bibitem[Skrutskie et al.(2006)]{skrutskie06} Skrutskie M.F., Cutri R.M., 
Stiening R., et al. 2006, AJ 131, 1163

\bibitem[Tody(1993)]{Tody1993} Tody D., 1993, in ASP Conf. 52,
Astronomical Data Analysis Software and Systems II, ed. R.J. Hanisch,
R.J.V. Brissenden, \& J. Barnes, p. 173

\bibitem[Ukwatta et al.(2010a)]{ubb10} Ukwatta T.N., Barthelmy S.D., 
  Baumgartner W.H., et al. 2010, GCN \#10870

\bibitem[Ukwatta et al.(2010b)]{use10} Ukwatta T.N., Stratta G., Evans P.A. et al.
 2010,  GCN report 191.1

\bibitem[Updike et al.(2010)]{unn10} Updike A., Nicuesa Guelbentu A., 
  Nardini M., Kr\"uhler T., Greiner J., 2010, GCN \#10874

\bibitem[Vlasis et al.(2011)]{vlasis11} Vlasis A., van Eerten H.J., Meliani Z.,
 Keppens R., 2011, MN 415, 279

\bibitem[Waxman \& Draine(2000)]{wad00} Waxman E., Draine B.T., 2000, 
ApJ 537, 796

\bibitem[Wijers et al.(1997)]{wij97} Wijers R.A.M.J. et al. 1997, MN 288, L51

\bibitem[Wijers \& Galama(1999)]{wig99} Wijers R.A.M.J.,  Galama T., 
1999, ApJ 523, 177

\bibitem[Yun \& Carilli(2002)]{yuc02} Yun M.S., Carilli C.L., 2002, ApJ 568, 88

\bibitem[Zauderer et al.(2012)]{zlb12} Zauderer B., Laskar T., Berger E., 
2012, GCN 13900

\bibitem[Zhang \& Meszaros(2004)]{zhm04} Zhang B.,  M{\'e}sz{\'a}ros P.,
2004, Int. J. Mod. Phys. A19, 2385

\end{thebibliography}
\end{document}